\documentclass[prl,twocolumn,showpacs,amsmath,amssymb,superscriptaddress]{revtex4-1}
\usepackage{epsfig,epstopdf} 
\usepackage{graphicx}
\usepackage{dcolumn}
\usepackage{bm}
\usepackage{bbm}
\usepackage{ulem}
\usepackage{color}
\usepackage{slashed}
\usepackage{hyperref}
\usepackage{mathrsfs}
\usepackage{subfigure}
\usepackage{verbatim}
\usepackage{dsfont}
\usepackage{float}
\usepackage{booktabs}
\usepackage{tikz}

\newcommand{\nc}{\newcommand}

\nc{\be}{\begin{equation}} \nc{\ee}{\end{equation}}
\nc{\bea}{\begin{eqnarray}} \nc{\eea}{\end{eqnarray}}
\nc{\bean}{\begin{eqnarray*}} \nc{\eean}{\end{eqnarray*}}
\nc{\dg}{\dagger}
\nc{\ua}{\uparrow} \nc{\da}{\downarrow}
\nc{\lag}{\langle} \nc{\rag}{\rangle}

\nc{\red}[1]{\textcolor{red}{#1}}

\begin{document}

\title{Higher Order Topology and Nodal Topological Superconductivity in Fe(Se,Te) Heterostructures}
\author{Rui-Xing Zhang}
\email{ruixing@umd.edu}
\affiliation{Condensed Matter Theory Center and Joint Quantum Institute, Department of Physics, University of Maryland, College Park, Maryland 20742-4111, USA}
\author{William S. Cole}
\affiliation{Condensed Matter Theory Center and Joint Quantum Institute, Department of Physics, University of Maryland, College Park, Maryland 20742-4111, USA}
\author{Xianxin Wu}
\affiliation{Institut f\"ur Theoretische Physik und Astrophysik, Universit\"at W\"urzburg, Am Hubland Campus S\"ud, W\"urzburg 97074, Germany}
\author{S. Das Sarma}
\affiliation{Condensed Matter Theory Center and Joint Quantum Institute, Department of Physics, University of Maryland, College Park, Maryland 20742-4111, USA}

\begin{abstract}
We show theoretically that a heterostructure of monolayer FeTe$_{1-x}$Se$_x$ - a superconducting quantum spin Hall material - with a monolayer of FeTe - a bicollinear antiferromagnet - realizes a higher order topological superconductor phase characterized by emergent Majorana zero modes pinned to the sample corners. We provide a minimal effective model for this system, analyze the origin of higher order topology, and fully characterize the topological phase diagram. Despite the conventional s-wave pairing, we find rather surprising emergence of a novel topological nodal superconductor in the phase diagram. Featured by edge-dependent Majorana flat bands, the topological nodal phase is protected by an antiferromagnetic chiral symmetry. We also discuss the experimental feasibility, the estimation of realistic model parameters, and the robustness of the Majorana corner modes against magnetic disorder. Our work provides a new experimentally feasible high-temperature platform for both higher order topology and non-Abelian Majorana physics.
\end{abstract}
\date{\today}
\maketitle

{\it Introduction} - For the past decade iron-based superconductors (FeSCs) have been a central research theme in condensed matter physics, owing to their high superconducting (SC) transition temperature $T_c$, rich phase diagrams, and in particular the puzzle of the origin of pairing \cite{kamihara2008iron,hanaguri2010unconventional,paglione2010high,wang2011electron,hirschfeld2011gap,chubukov2012pairing,chubukov2014fe}. While the underlying microscopic mechanisms of SC in both bulk and monolayer FeSCs remain controversial, remarkable progress has been made recently towards revealing their nontrivial topological properties \cite{hao2014topological,wu2015cafeas,wu2016topological,xu2016topological,zhang2018observation,wang2018evidence,zhang2018multiple,hao2018topological,machida2018zero}. As the prototypical example of topological FeSC, bulk FeTe$_{1-x}$Se$_x$ (FTS) with $x=0.45$ hosts a helical Dirac surface state above $T_c$, as confirmed by angle-resolved photoemission spectroscopy (ARPES) measurements \cite{zhang2018observation,zhang2018multiple}. Below $T_c$, strong evidence for Majorana vortex bound states has been found reproducibly in several scanning tunneling microscopy (STM) experiments \cite{wang2018evidence,machida2018zero,kong2019observation,zhu2019observation}, following from the theoretical prediction of surface topological superconductivity developed via a ``self-proximity" effect \cite{fu2008superconducting}. Similar to its bulk counterpart, the normal-state band structure of monolayer FTS has been theoretically predicted to be topological \cite{wu2016topological}. This prediction is further supported by a recent systematic ARPES measurement of monolayer FTS \cite{shi2017fete1,peng2019observation}, clearly revealing a bulk topological phase transition (band gap closing at $\Gamma$) by continuously changing the value of $x$. With the highest $T_c$ among FeSCs \cite{qing2012interface,wen2014direct,ge2015superconductivity}, one might wonder whether FTS monolayer also offers a new high temperature platform for topological Majorana physics. It should be noted, however, that the coexistence of nontrivial band topology and SC does not {\it guarantee} topological superconductivity (TSC). In fact, two-dimensional (2d), time reversal invariant TSC requires very strict conditions for both the Fermi surface geometry and SC pairing symmetry \cite{qi2009time,zhang2013time}. With the puzzle of pairing symmetry unresolved \cite{huang2017monolayer}, the question of TSC in monolayer FTS remains open, although the answer is very likely negative.

\begin{figure}[t]
	\centering
	\includegraphics[width=0.5\textwidth]{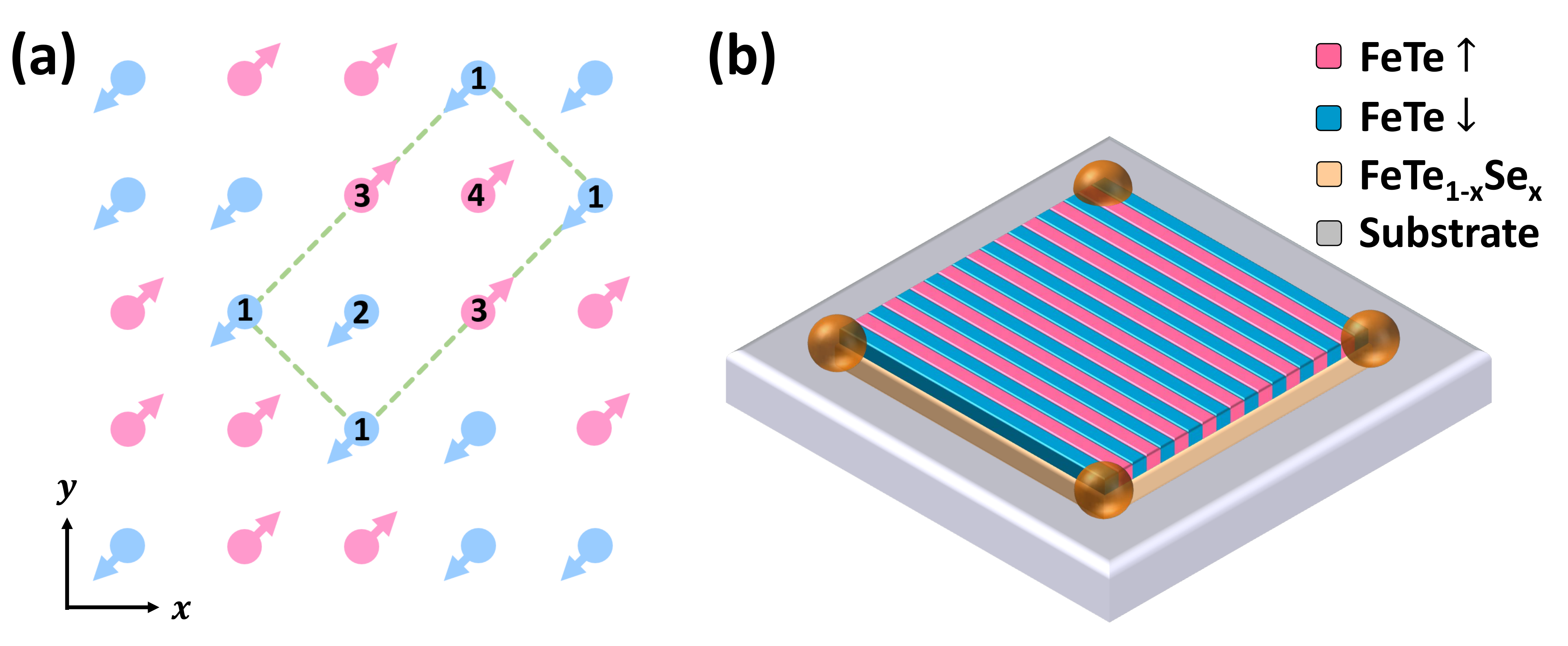}
	\caption{(a) Schematic plot of bicollinear antiferromagnetic order in FeTe. The circle and its arrow represent the Fe atom and its magnetic moment. (b) Schematic plot of the FTS/FeTe heterostructure with corner-localized Majorana modes.}
	\label{Fig: schematic}
\end{figure}

In this Letter, we provide an alternative pairing-symmetry-independent route to obtain Majorana bound states in monolayer FTS systems. We demonstrate that Majorana zero modes emerge at physical corners of a sample when a FeTe layer is deposited on top of the FTS monolayer, as shown in Fig. \ref{Fig: schematic} (b). The bicollinear antiferromagentic order of FeTe \cite{ma2009first,bao2009tunable,li2009first,manna2017interfacial} is the key enabling higher order topology \cite{benalcazar2017quantized,benalcazar2017electric,zhang2013surface,schindler2018higher,langbehn2017reflection,khalaf2018higher,liu2018majorana,peng2018proximity,shapourian2018topological,volpez2019second,wang2018high,yan2018majorana,wang2018weak,wu2019higher,zhu2018tunable,bultinck2019three,pan2018lattice,ghorashi2019second,franca2019phase,hsu2018majorana,kudo2019higher,yan2019higher} in this heterostructure binding localized Majorana zero modes, without relying on the choice of pairing symmetry. This heterostructure-based mechanism is essentially different from our earlier proposal for bulk FTS, where the higher order topology is enabled by unconventional $s_{\pm}$ pairing \cite{zhang2019helical}. We construct a minimal lattice model to explain the origin of higher order topology in this heterostructure and we also study the stability of Majorana corner modes with respect to finite chemical potential $\mu$ and disorder effects. In the large $\mu$ limit, the FeTe layer also enables a novel topological nodal SC phase with symmetry protected edge Majorana flat bands, even when the SC pairing is singlet s-wave.   

{\it Model Hamiltonian} - The low-energy theory of monolayer FTS around the $\Gamma$-point is a superconducting version of the Bernevig-Hughes-Zhang (BHZ) model \cite{bernevig2006quantum,wu2016topological}. While the pairing mechanism in monolayer iron chalcogenide systems is still under debate, a conventional s-wave singlet pairing $\Delta$ will suffice for our purpose. The Hamiltonian for FTS is then
\bea
h_{\rm BHZ} &=&
[(m-4B)+2B(\cos k_x + \cos k_y)]\Gamma_5 - \mu \Gamma_{67}  \nonumber \\
&&+ A (\sin k_x\Gamma_{167} - \sin k_y \Gamma_2)  + \Delta \Gamma_{137}
\label{Eq: BHZ Hamiltonian}
\eea
in terms of a choice of $8 \times 8$ $\Gamma$ matrices 
\bea
\Gamma_1 &=& \tau_z \otimes \sigma_x \otimes s_z,\ 
\Gamma_2 = \tau_z \otimes \sigma_y \otimes s_0 \nonumber \\
\Gamma_3 &=& \tau_z \otimes \sigma_x \otimes s_x,\ 
\Gamma_4 = \tau_z \otimes \sigma_x \otimes s_y \nonumber \\
\Gamma_5 &=& \tau_z \otimes \sigma_z \otimes s_0,\ 
\Gamma_6 = \tau_x \otimes \sigma_0 \otimes s_0 \nonumber \\
\Gamma_7 &=& \tau_y \otimes \sigma_0 \otimes s_0,
\eea
with $\Gamma_{jk} = \Gamma_j\Gamma_k/i$ and $\Gamma_{jkl} = \Gamma_j\Gamma_k\Gamma_l/i$ for $j\neq k \neq l \in \{1,2,...,8\}$.
Here $s$, $\sigma$ and $\tau$ Pauli matrices denote spin, orbital and particle-hole degrees of freedom, respectively. When $0<m<8B$, the normal-state part of $h_{\rm BHZ}$ is  topologically nontrivial and possesses helical edge modes. However, the s-wave SC pairing necessarily trivializes the band topology of the full BdG model and introduces a pairing gap on the edge.

Covering the FTS monolayer with a monolayer FeTe introduces an exchange coupling with the bicollinear AFM order of FeTe to the system. Unlike a conventional collinear AFM, the magnetic moments in the bicollinear AFM flip their orientation every two atoms along the diagonal direction (e.g. the [11] direction) \cite{ma2009first}, as shown in Fig. \ref{Fig: schematic} (a). As a result, the unit cell is enlarged to contain four inequivalent atoms, labeled by a sublattice index $i=1,2,3,4$. The new unit cell is characterized by the lattice vectors $\tilde{\bf a}_x = 2 ({\bf a}_x + {\bf a}_y), \tilde{\bf a}_y = -{\bf a}_x + {\bf a}_y$, where ${\bf a}_{x,y}$ are the lattice vectors of the original square lattice BHZ model. $k_{\tilde{x}}$ and $k_{\tilde{y}}$ denote the crystal momenta in the folded Brillouin zone and we will use $M$ to represent the interlayer exchange coupling between FTS and FT. The matrix form of the full Hamiltonian $H_\text{FTS}$ with both SC and AFM is shown in the Supplemental Material \cite{supplementary}.

\begin{figure*}[t]
	\centering
	\includegraphics[width=0.85\textwidth]{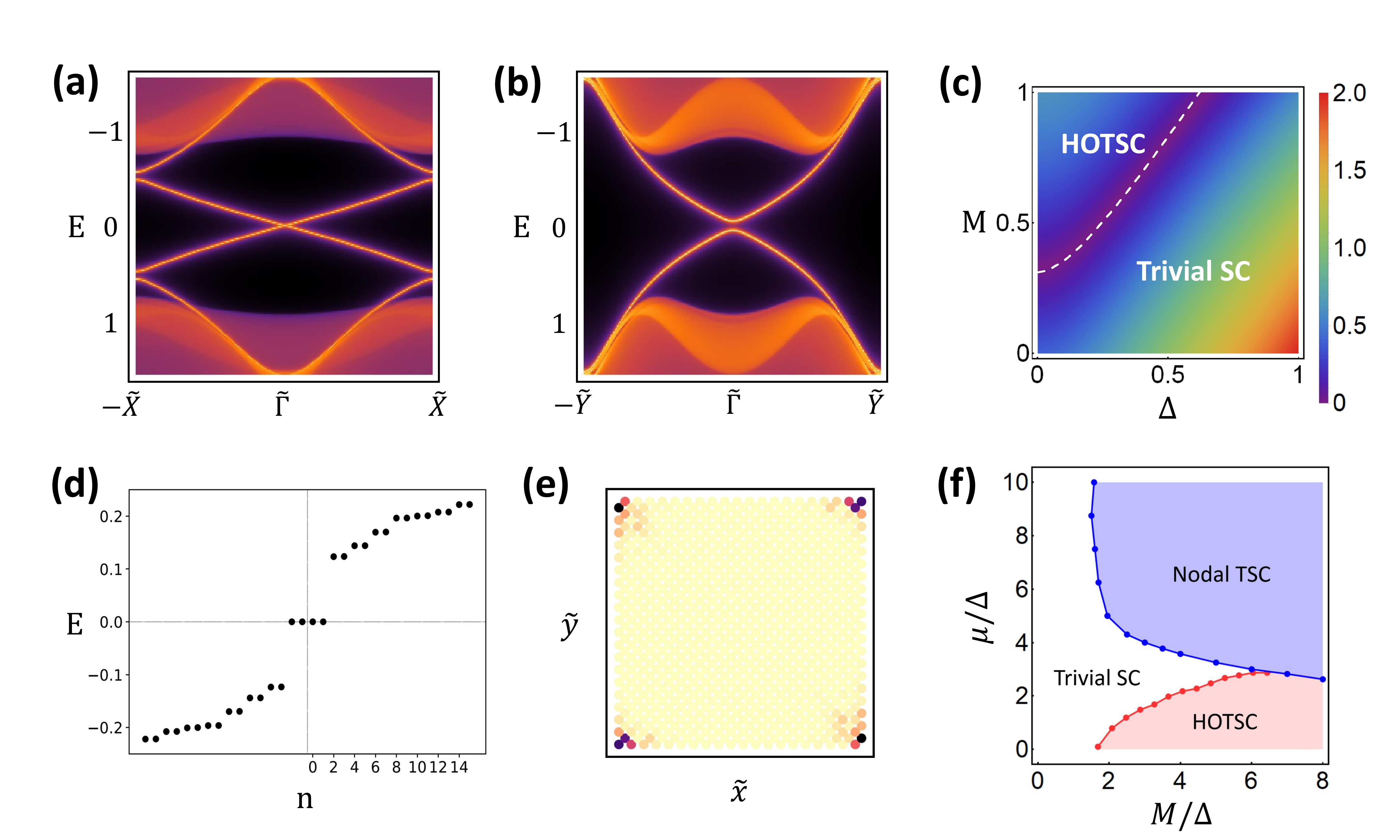}
	\caption{With $(A,B,m,M,\Delta)=(1,1,2,0.2,0)$, dispersions of $\tilde{x}$-edge (AFM) and $\tilde{y}$-edge (FM) are plotted in (a) and (b), respectively. (c) Topological phase diagram for a fixed $\mu=0.2$. The white dashed line shows the analytical results of Eq. \ref{Eq: Topological criterion}. (d) Energy spectrum of $H_\text{FTS}$ with open boundary conditions in both $\tilde{x}$ and $\tilde{y}$ directions, which clearly reveals four Majorana zero modes. (e) Spatial profile of the Majorana zero modes in (d). (f) Topological phase diagram with respect to $M$ and $\mu$ at a fixed $\Delta$.}
	\label{Fig: Cornermode}
\end{figure*}

{\it Majorana Corner Modes} -  To understand the emergence of topological Majorana zero modes in our system, it is instructive to first switch off superconductivity and study the topological consequence of the bicollinear AFM. Despite explicitly breaking the time reversal symmetry $\Theta$, introducing AFM to a QSH system does NOT necessarily destroy the helical edge states. These edge states are instead now protected by an effective TRS $\Theta_M = \Theta e^{ik_{\tilde{x}}/2}$, which combines $\Theta$ with a half-unit-cell translation along ${\bf a}_{\tilde{x}}$. As shown in Fig. \ref{Fig: schematic} (a), $\Theta_M$ swaps electrons of index $i=1,2$ with those of $i=3,4$. Therefore, $\Theta_M$ is a magnetic space group operation and has a crucial difference from the conventional TRS $\Theta$ \cite{mong2010antiferromagnetic,liu2013antiferromagnetic,zhang2015topological}: the Kramers degeneracy of $\Theta_M$ only arises at the high symmetry points with $k_{\tilde{x}}=0$. 

Due to the crystalline nature of $\Theta_M$, not every edge preserves $\Theta_M$ and is capable of hosting helical edge states. In particular, as shown in Fig. \ref{Fig: schematic} (a), the magnetic configuration of the $\tilde{\bf y}$ edge (which is parallel to $\tilde{\bf a}_y$) is ferromagnetic (FM), which locally breaks $\Theta_M$ and produces an edge magnetic gap. This is in contrast to gapless edges (such as the $\tilde{\bf x}$ edge) with AFM ordering and $\Theta_M$ protection. To verify this picture, we have used the iterative Green function method to numerically calculate the edge dispersion with finite exchange coupling $M$ and zero $\Delta$ for both $\tilde{\bf x}$ and $\tilde{\bf y}$ edges. As shown in Fig. \ref{Fig: Cornermode} (a), the $\tilde{\bf x}$ edge has a Kramers degeneracy at $\tilde{\Gamma}$ (with $\Theta_M^2 = -1$) but not at $\tilde{X}$ (with $\Theta_M^2 = +1$). Meanwhile, Fig. \ref{Fig: Cornermode} (b) clearly shows the magnetic gap at $\tilde{\Gamma}$ on the $\tilde{\bf y}$ edge, which follows our expectation.

Now we include SC in our discussion. Through the self-proximity effect accompanying the development of bulk SC in the FTS layer, the gapless $\tilde{\bf x}$ edge opens a SC gap. For the $\tilde{\bf y}$ edge, however, there exists a competition between the edge FM gap and the edge SC gap. In particular, when the FM gap dominates the $\tilde{\bf x}$ edge, the corner between $\tilde{\bf x}$ and $\tilde{\bf y}$ edges represents a zero-dimensional domain wall between SC and FM gaps, which necessarily binds a single Majorana zero mode to the corner \cite{fu2009josephson,alicea2012new}, thus enabling higher order topology. 

Therefore, the higher order topology in the heterostructure is controlled by the character of the $\tilde{\bf y}$ edge gap where SC and FM compete. This motivates us to construct an effective theory of the $\tilde{\bf y}$ edge that describes the competition between FM and SC,
\bea
h_{\tilde{\bf y}} &=& k_{\tilde{y}} \tau_0 \otimes \sigma_z + \delta_M \tau_z \otimes \sigma_x + \Delta \tau_y\otimes \sigma_y - \mu \tau_z \otimes \sigma_0. \nonumber \\
&&
\label{Eq: Edge Theory}
\eea
Here $\delta_M$ is the effective exchange coupling on the $\tilde{\bf y}$ edge, which orignates from the edge projection of the bulk AFM order $M$. While it is generally difficult to analytically express $\delta_M$ in terms of $M$, we numerically confirm a simple linear relation with $\delta_M \approx \beta_M M $. The linear coefficient $\beta_M$ depends on the details of hopping parameters and we find $\beta_M \simeq 0.678$ for our choice of parameters. For nonzero $\delta_M$ and $\Delta$, the edge topological phase transition occurs when the energy gap of $h_{\tilde{\bf y}}$ closes. The topological condition of HOTSC is thus given by \cite{supplementary},
\bea
M^2 &>& \frac{1}{\beta_M^2}(\mu^2 + \Delta^2)
\label{Eq: Topological criterion}
\eea
when FM exceeds SC on the $\tilde{\bf y}$ edge. To further confirm Eq. \ref{Eq: Topological criterion}, we numerically map out the energy gap distribution in the parameter space spanned by $M$ and $\Delta$ at a fixed $\mu$. As shown in Fig. \ref{Fig: Cornermode} (c), the topological phase transition predicted by Eq. \ref{Eq: Topological criterion} (white dashed line) agrees well with the colormap of the edge gap from a numerical nanoribbon calculation (where the gap closing regions are labeled in purple). 

Following this topological criterion, we calculate the eigenvalues of $H_{\rm FTS}$ on a $20\tilde{a}_y \times 10\tilde{a}_x$ open cluster by direct diagonalization. As shown in Fig. \ref{Fig: Cornermode} (d), in the topological phase at $(M, \Delta, \mu) = (0.6, 0.2, 0.0)$, four Majorana zero modes are found to live inside the edge gap. We plot the combined spatial profile of these four zero modes in Fig. \ref{Fig: Cornermode} (e), and additionally confirm that they are exponentially localized at the corners of the system. These corner localized 0d Majorana bound states are the hallmark of higher order topology in this 2d system.

{\it Emergent Nodal TSC} - In the small $\mu$ limit, the edge topological condition in Eq. \ref{Eq: Topological criterion} provides a simple analytical diagnostic for the appearance of Majorana corner modes. To explore the fate of higher order topology at finite $\mu$, we fix the value of $\Delta$ and numerically map out the topological phase diagram tuning $M$ and $\mu$. In Fig. \ref{Fig: Cornermode} (f), the HOTSC phase and the trivial phase are denoted by the red and white regions respectively. The phase boundary that separates the HOTSC and trivial SC corresponds to the gap closing of FM edge. In addition, an emergent nodal superconducting phase (blue region) is found to dominate the phase diagram when both $\mu$ and $M$ are large. Since only conventional s-wave singlet pairing is considered in our model, this nodal structure is unusual and emerges from the combined effects of AFM and SC. 

\begin{figure}[t]
	\centering
	\includegraphics[width=0.48\textwidth]{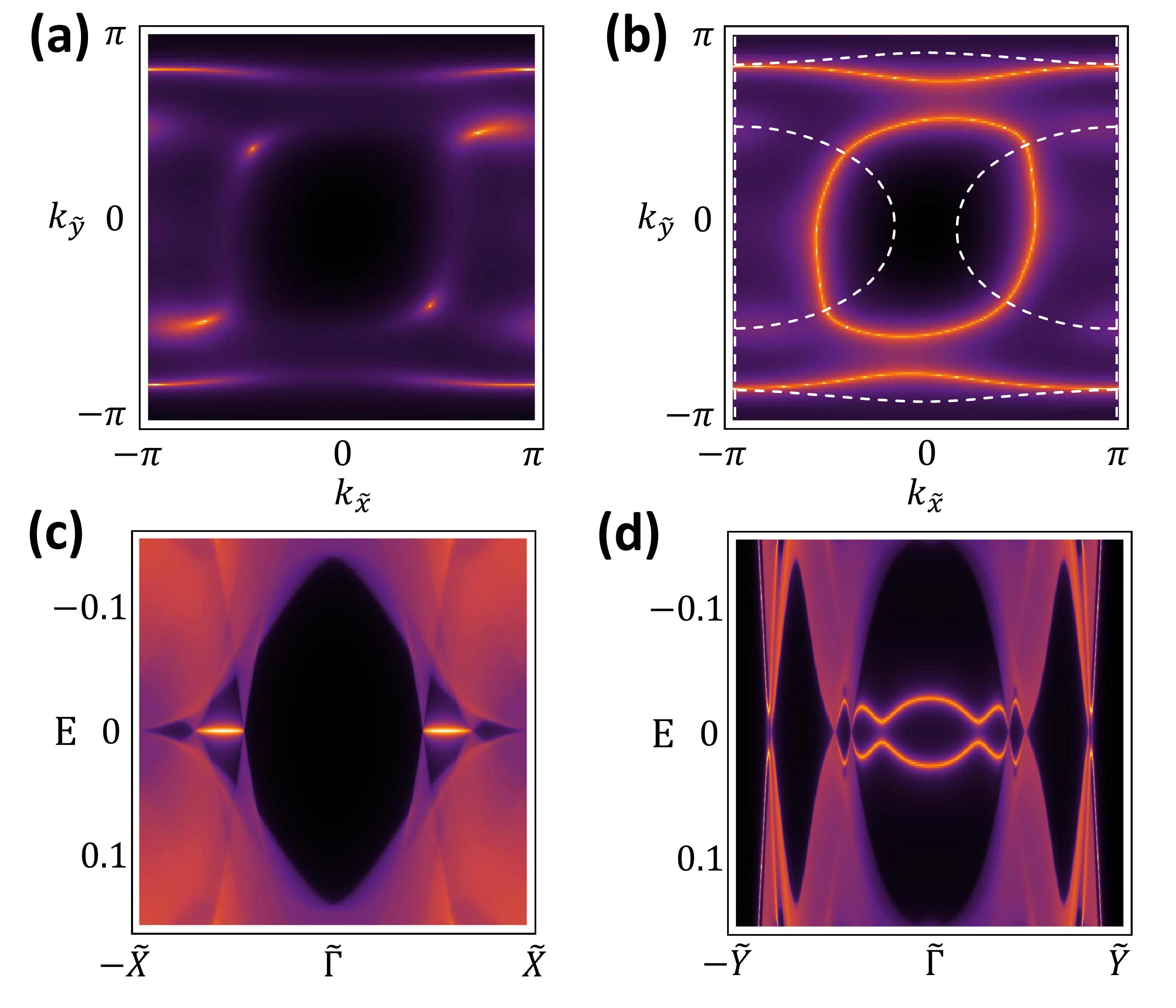}
	\caption{(a) Position of SC nodes in the Brillouin zone for the nodal SC phase. (b) Fermi surface of the corresponding normal band structure and the zero-pairing contours (white dashed lines). (c) and (d) show the dispersions of the AFM edge and the FM edge for the nodal SC phase, respectively. The AFM edge hosts symmetry protected Majorana flat bands.}
	\label{Fig: Nodal}
\end{figure}

The origin of the emergent nodal SC can be understood by projecting s-wave pairing onto the bulk Fermi surface. At zero $M$, the effective pairing gap $\Delta_{FS}({\bf k})$ is always uniform on the Fermi surface, which simply signals the uniform, isotropic s-wave pairing. As $M$ is turned on from zero, the Fermi surface develops a spin texture such that $\Delta_{FS}({\bf k})$ becomes anisotropic in the Brillouin zone and thus develops momentum contours with $\Delta_{FS}({\bf k})=0$. For example, the Fermi surface for $(M,\Delta,\mu)=(0.6,0.2,1.2)$ is mapped out in Fig. \ref{Fig: Nodal} (a), which clearly shows the position of SC nodes in the spectrum. As a comparison, we plot the Fermi surface of the normal band structure alone in Fig. \ref{Fig: Nodal} (b), along with the calculated zero-pairing contours (white dashed lines). As expected, the BdG spectrum has nodal points where the zero-pairing contour intersects the normal state Fermi surface.

These emergent bulk SC nodes carry non-trivial topological charge and thus lead to interesting boundary Majorana physics. By combining the effective TRS $\Theta_M$ and particle-hole symmetry $\Pi$, an AFM chiral symmetry operation is defined as
\bea
{\cal C} = \Theta_M \Pi,
\eea  
which anticommutes with $H_\text{FTS}$. The AFM chiral symmetry allows us in turn to define a topological charge $Q$ \cite{schnyder2011topological,yu2018singlet,supplementary}, and we numerically find
\bea
|Q|=2,
\eea
for every SC node.

The bulk-boundary correspondence then implies the existence of edge Majorana flat bands between SC nodes with opposite topological charges in a nanoribbon geometry. In Fig. \ref{Fig: Nodal} (c) and (d), we show the calculated edge spectrum for the AFM $\tilde{y}$ edge and FM $\tilde{x}$ edge, respectively. As expected, the $\tilde{y}$ edge hosts zero-energy Majorana flat bands between the projections of the nodal points. These Majorana flat bands are doubly degenerate due to $|Q|=2$. On the $\tilde{x}$ edge, however, the AFM chiral symmetry is explicitly broken because of the absence of $\Theta_M$. Therefore, the inter-node edge modes are not protected by ${\cal C}$ and therefore need not be pinned to zero energy. As expected, the edge modes in Fig. \ref{Fig: Nodal} (d) are found to hybridize with each other with a nonzero splitting which shifts the modes from zero energy. The edge-dependent Majorana flat bands are a unique feature of the AFM chiral symmetry-protected nodal TSC phase.

{\it Feasibility of Experimental Realization} - We now discuss the experimental feasibility of our proposal. We first notice that the fabrication techniques for iron chalcogenide heterostructures are well-developed \cite{wen2014direct,sun2014high,nabeshima2017growth}. In particular, bilayers of different iron chalcogenide layers (for example, a FeSe layer and a FeTe layer) were found to be coherently constrained to each other \cite{nabeshima2017growth}, which should hold in our proposed FeTe$_{1-x}$Se$_{x}$/FeTe bilayer as well. The precise epitaxial lattice matching between different iron chalcogenide layers greatly facilitates the edge characterization and identification of corner Majorana signal in realistic materials.

We also attempt to make some realistic estimates on the energy scales of physical quantities involved in the topological condition of Eq. \ref{Eq: Topological criterion}. ARPES studies on the monolayer FeSe system reveal a SC gap $\Delta$ of about 10 meV \cite{zhang2016superconducting}. The magnetic structure of FeTe has been measured in Ref. \cite{bao2009tunable,li2009first}, which leads to local magnetic moments about $1.65 \mu_B$ along b axis (parallel spin axis), where $\mu_B$ is the Bohr magneton. The local magnetic moment of FeTe layer induces a magnetic proximity effect through exchange coupling with the FTS layer. To evaluate the scale of the induced exchange coupling, we perform a DFT calculation of a bilayer FeSe system, introducing ferromagnetism to the top FeSe layer \cite{supplementary}. With a magnetic moment of $2.76 \mu_B$, the proximity-induced exchange coupling of the bottom layer is around 100 meV. Thus, for the experimentally observed magnetic moment of $1.65 \mu_B$ in FeTe, the induced exchange coupling in the FTS layer is expected to be $M\approx 60$ meV. Given that $\beta_M\sim 0.5$, the edge FM potential is still much greater than the edge SC potential and thus the topological condition for Majorana corner modes is always satisfied for a small chemical potential.

{\it Conclusion and Discussion} - 
We have established the higher order TSC phase with Majorana corner modes in monolayer Fe(Se,Te) heterostructures, together with an emergent symmetry-protected nodal TSC phase. 
In addition to our proposed heterostructure being feasible to create, both the Majorana corner modes and the dispersing Majorana edge flat bands exhibit distinct features in local spectroscopy and should therefore be experimentally visible using standard STM techniques. Our proposed realization of Majorana corner modes does not rely on fine-tuning of the chemical potential, nor does it require perfect ordering of moments in the FeTe layer. In the Supplemental Material \cite{supplementary}, we consider a disordered model with a fixed density of magnetic defects where $\pm M \rightarrow \mp M$, which explicitly breaks $\Theta_M$, and we find that the corner modes persist even for a substantial defect density. Finally, unlike most previous proposals for higher order TSC \cite{yan2018majorana,wang2018high,wang2018weak,zhang2019helical}, our setup does not require an unconventional pairing symmetry. Rather, the combination of AFM and conventional s-wave pairing effectively mimics anisotropic pairing leading to generic higher order topological SC. However, it would be interesting (and possibly experimentally relevant for FTS) to generalize to unconventional pairing. In fact, the full landscape of trivial vs. topological normal state band structure, uniform vs. nonuniform magnetism, and conventional vs. unconventional pairing symmetry appears to be quite rich and already at hand in iron-based materials.

{\it Acknowledgment} - R.-X.Z is indebted to Chao-Xing Liu, Jiabin Yu, Fengcheng Wu and Biao Lian for helpful discussions. This work is supported by Laboratory for Physical Sciences and Microsoft. R.-X.Z is supported by a JQI Postdoctoral Fellowship.

We have recently become aware of a related work on the monolayer FTS, where the higher order topology is driven by an in-plane magnetic field \cite{wu2019high}.

\bibliographystyle{apsrev4-1}
\bibliography{corner}

\begin{thebibliography}{67}%
\makeatletter
\providecommand \@ifxundefined [1]{%
 \@ifx{#1\undefined}
}%
\providecommand \@ifnum [1]{%
 \ifnum #1\expandafter \@firstoftwo
 \else \expandafter \@secondoftwo
 \fi
}%
\providecommand \@ifx [1]{%
 \ifx #1\expandafter \@firstoftwo
 \else \expandafter \@secondoftwo
 \fi
}%
\providecommand \natexlab [1]{#1}%
\providecommand \enquote  [1]{``#1''}%
\providecommand \bibnamefont  [1]{#1}%
\providecommand \bibfnamefont [1]{#1}%
\providecommand \citenamefont [1]{#1}%
\providecommand \href@noop [0]{\@secondoftwo}%
\providecommand \href [0]{\begingroup \@sanitize@url \@href}%
\providecommand \@href[1]{\@@startlink{#1}\@@href}%
\providecommand \@@href[1]{\endgroup#1\@@endlink}%
\providecommand \@sanitize@url [0]{\catcode `\\12\catcode `\$12\catcode
  `\&12\catcode `\#12\catcode `\^12\catcode `\_12\catcode `\%12\relax}%
\providecommand \@@startlink[1]{}%
\providecommand \@@endlink[0]{}%
\providecommand \url  [0]{\begingroup\@sanitize@url \@url }%
\providecommand \@url [1]{\endgroup\@href {#1}{\urlprefix }}%
\providecommand \urlprefix  [0]{URL }%
\providecommand \Eprint [0]{\href }%
\providecommand \doibase [0]{http://dx.doi.org/}%
\providecommand \selectlanguage [0]{\@gobble}%
\providecommand \bibinfo  [0]{\@secondoftwo}%
\providecommand \bibfield  [0]{\@secondoftwo}%
\providecommand \translation [1]{[#1]}%
\providecommand \BibitemOpen [0]{}%
\providecommand \bibitemStop [0]{}%
\providecommand \bibitemNoStop [0]{.\EOS\space}%
\providecommand \EOS [0]{\spacefactor3000\relax}%
\providecommand \BibitemShut  [1]{\csname bibitem#1\endcsname}%
\let\auto@bib@innerbib\@empty
\bibitem [{\citenamefont {Kamihara}\ \emph {et~al.}(2008)\citenamefont
  {Kamihara}, \citenamefont {Watanabe}, \citenamefont {Hirano},\ and\
  \citenamefont {Hosono}}]{kamihara2008iron}%
  \BibitemOpen
  \bibfield  {author} {\bibinfo {author} {\bibfnamefont {Y.}~\bibnamefont
  {Kamihara}}, \bibinfo {author} {\bibfnamefont {T.}~\bibnamefont {Watanabe}},
  \bibinfo {author} {\bibfnamefont {M.}~\bibnamefont {Hirano}}, \ and\ \bibinfo
  {author} {\bibfnamefont {H.}~\bibnamefont {Hosono}},\ }\href {\doibase
  10.1021/ja800073m} {\bibfield  {journal} {\bibinfo  {journal} {Journal of the
  American Chemical Society}\ }\textbf {\bibinfo {volume} {130}},\ \bibinfo
  {pages} {3296} (\bibinfo {year} {2008})}\BibitemShut {NoStop}%
\bibitem [{\citenamefont {Hanaguri}\ \emph {et~al.}(2010)\citenamefont
  {Hanaguri}, \citenamefont {Niitaka}, \citenamefont {Kuroki},\ and\
  \citenamefont {Takagi}}]{hanaguri2010unconventional}%
  \BibitemOpen
  \bibfield  {author} {\bibinfo {author} {\bibfnamefont {T.}~\bibnamefont
  {Hanaguri}}, \bibinfo {author} {\bibfnamefont {S.}~\bibnamefont {Niitaka}},
  \bibinfo {author} {\bibfnamefont {K.}~\bibnamefont {Kuroki}}, \ and\ \bibinfo
  {author} {\bibfnamefont {H.}~\bibnamefont {Takagi}},\ }\href {\doibase
  10.1126/science.1187399} {\bibfield  {journal} {\bibinfo  {journal}
  {Science}\ }\textbf {\bibinfo {volume} {328}},\ \bibinfo {pages} {474}
  (\bibinfo {year} {2010})}\BibitemShut {NoStop}%
\bibitem [{\citenamefont {{Paglione}}\ and\ \citenamefont
  {{Greene}}(2010)}]{paglione2010high}%
  \BibitemOpen
  \bibfield  {author} {\bibinfo {author} {\bibfnamefont {J.}~\bibnamefont
  {{Paglione}}}\ and\ \bibinfo {author} {\bibfnamefont {R.~L.}\ \bibnamefont
  {{Greene}}},\ }\href {\doibase 10.1038/nphys1759} {\bibfield  {journal}
  {\bibinfo  {journal} {Nature Physics}\ }\textbf {\bibinfo {volume} {6}},\
  \bibinfo {pages} {645} (\bibinfo {year} {2010})}\BibitemShut {NoStop}%
\bibitem [{\citenamefont {Wang}\ and\ \citenamefont
  {Lee}(2011)}]{wang2011electron}%
  \BibitemOpen
  \bibfield  {author} {\bibinfo {author} {\bibfnamefont {F.}~\bibnamefont
  {Wang}}\ and\ \bibinfo {author} {\bibfnamefont {D.-H.}\ \bibnamefont {Lee}},\
  }\href {\doibase 10.1126/science.1200182} {\bibfield  {journal} {\bibinfo
  {journal} {Science}\ }\textbf {\bibinfo {volume} {332}},\ \bibinfo {pages}
  {200} (\bibinfo {year} {2011})}\BibitemShut {NoStop}%
\bibitem [{\citenamefont {{Hirschfeld}}\ \emph {et~al.}(2011)\citenamefont
  {{Hirschfeld}}, \citenamefont {{Korshunov}},\ and\ \citenamefont
  {{Mazin}}}]{hirschfeld2011gap}%
  \BibitemOpen
  \bibfield  {author} {\bibinfo {author} {\bibfnamefont {P.~J.}\ \bibnamefont
  {{Hirschfeld}}}, \bibinfo {author} {\bibfnamefont {M.~M.}\ \bibnamefont
  {{Korshunov}}}, \ and\ \bibinfo {author} {\bibfnamefont {I.~I.}\ \bibnamefont
  {{Mazin}}},\ }\href {\doibase 10.1088/0034-4885/74/12/124508} {\bibfield
  {journal} {\bibinfo  {journal} {Reports on Progress in Physics}\ }\textbf
  {\bibinfo {volume} {74}},\ \bibinfo {eid} {124508} (\bibinfo {year}
  {2011})}\BibitemShut {NoStop}%
\bibitem [{\citenamefont {Chubukov}(2012)}]{chubukov2012pairing}%
  \BibitemOpen
  \bibfield  {author} {\bibinfo {author} {\bibfnamefont {A.}~\bibnamefont
  {Chubukov}},\ }\href {\doibase 10.1146/annurev-conmatphys-020911-125055}
  {\bibfield  {journal} {\bibinfo  {journal} {Annual Review of Condensed Matter
  Physics}\ }\textbf {\bibinfo {volume} {3}},\ \bibinfo {pages} {57} (\bibinfo
  {year} {2012})}\BibitemShut {NoStop}%
\bibitem [{\citenamefont {{Chubukov}}\ and\ \citenamefont
  {{Hirschfeld}}(2015)}]{chubukov2014fe}%
  \BibitemOpen
  \bibfield  {author} {\bibinfo {author} {\bibfnamefont {A.}~\bibnamefont
  {{Chubukov}}}\ and\ \bibinfo {author} {\bibfnamefont {P.~J.}\ \bibnamefont
  {{Hirschfeld}}},\ }\href {\doibase 10.1063/PT.3.2818} {\bibfield  {journal}
  {\bibinfo  {journal} {Physics Today}\ }\textbf {\bibinfo {volume} {68}},\
  \bibinfo {pages} {46} (\bibinfo {year} {2015})}\BibitemShut {NoStop}%
\bibitem [{\citenamefont {Hao}\ and\ \citenamefont
  {Hu}(2014)}]{hao2014topological}%
  \BibitemOpen
  \bibfield  {author} {\bibinfo {author} {\bibfnamefont {N.}~\bibnamefont
  {Hao}}\ and\ \bibinfo {author} {\bibfnamefont {J.}~\bibnamefont {Hu}},\
  }\href {\doibase 10.1103/PhysRevX.4.031053} {\bibfield  {journal} {\bibinfo
  {journal} {Phys. Rev. X}\ }\textbf {\bibinfo {volume} {4}},\ \bibinfo {pages}
  {031053} (\bibinfo {year} {2014})}\BibitemShut {NoStop}%
\bibitem [{\citenamefont {Wu}\ \emph {et~al.}(2015)\citenamefont {Wu},
  \citenamefont {Qin}, \citenamefont {Liang}, \citenamefont {Le}, \citenamefont
  {Fan},\ and\ \citenamefont {Hu}}]{wu2015cafeas}%
  \BibitemOpen
  \bibfield  {author} {\bibinfo {author} {\bibfnamefont {X.}~\bibnamefont
  {Wu}}, \bibinfo {author} {\bibfnamefont {S.}~\bibnamefont {Qin}}, \bibinfo
  {author} {\bibfnamefont {Y.}~\bibnamefont {Liang}}, \bibinfo {author}
  {\bibfnamefont {C.}~\bibnamefont {Le}}, \bibinfo {author} {\bibfnamefont
  {H.}~\bibnamefont {Fan}}, \ and\ \bibinfo {author} {\bibfnamefont
  {J.}~\bibnamefont {Hu}},\ }\href {\doibase 10.1103/PhysRevB.91.081111}
  {\bibfield  {journal} {\bibinfo  {journal} {Phys. Rev. B}\ }\textbf {\bibinfo
  {volume} {91}},\ \bibinfo {pages} {081111} (\bibinfo {year}
  {2015})}\BibitemShut {NoStop}%
\bibitem [{\citenamefont {Wu}\ \emph {et~al.}(2016)\citenamefont {Wu},
  \citenamefont {Qin}, \citenamefont {Liang}, \citenamefont {Fan},\ and\
  \citenamefont {Hu}}]{wu2016topological}%
  \BibitemOpen
  \bibfield  {author} {\bibinfo {author} {\bibfnamefont {X.}~\bibnamefont
  {Wu}}, \bibinfo {author} {\bibfnamefont {S.}~\bibnamefont {Qin}}, \bibinfo
  {author} {\bibfnamefont {Y.}~\bibnamefont {Liang}}, \bibinfo {author}
  {\bibfnamefont {H.}~\bibnamefont {Fan}}, \ and\ \bibinfo {author}
  {\bibfnamefont {J.}~\bibnamefont {Hu}},\ }\href {\doibase
  10.1103/PhysRevB.93.115129} {\bibfield  {journal} {\bibinfo  {journal} {Phys.
  Rev. B}\ }\textbf {\bibinfo {volume} {93}},\ \bibinfo {pages} {115129}
  (\bibinfo {year} {2016})}\BibitemShut {NoStop}%
\bibitem [{\citenamefont {Xu}\ \emph {et~al.}(2016)\citenamefont {Xu},
  \citenamefont {Lian}, \citenamefont {Tang}, \citenamefont {Qi},\ and\
  \citenamefont {Zhang}}]{xu2016topological}%
  \BibitemOpen
  \bibfield  {author} {\bibinfo {author} {\bibfnamefont {G.}~\bibnamefont
  {Xu}}, \bibinfo {author} {\bibfnamefont {B.}~\bibnamefont {Lian}}, \bibinfo
  {author} {\bibfnamefont {P.}~\bibnamefont {Tang}}, \bibinfo {author}
  {\bibfnamefont {X.-L.}\ \bibnamefont {Qi}}, \ and\ \bibinfo {author}
  {\bibfnamefont {S.-C.}\ \bibnamefont {Zhang}},\ }\href {\doibase
  10.1103/PhysRevLett.117.047001} {\bibfield  {journal} {\bibinfo  {journal}
  {Phys. Rev. Lett.}\ }\textbf {\bibinfo {volume} {117}},\ \bibinfo {pages}
  {047001} (\bibinfo {year} {2016})}\BibitemShut {NoStop}%
\bibitem [{\citenamefont {{Zhang}}\ \emph {et~al.}(2018)\citenamefont
  {{Zhang}}, \citenamefont {{Yaji}}, \citenamefont {{Hashimoto}}, \citenamefont
  {{Ota}}, \citenamefont {{Kondo}}, \citenamefont {{Okazaki}}, \citenamefont
  {{Wang}}, \citenamefont {{Wen}}, \citenamefont {{Gu}}, \citenamefont
  {{Ding}},\ and\ \citenamefont {{Shin}}}]{zhang2018observation}%
  \BibitemOpen
  \bibfield  {author} {\bibinfo {author} {\bibfnamefont {P.}~\bibnamefont
  {{Zhang}}}, \bibinfo {author} {\bibfnamefont {K.}~\bibnamefont {{Yaji}}},
  \bibinfo {author} {\bibfnamefont {T.}~\bibnamefont {{Hashimoto}}}, \bibinfo
  {author} {\bibfnamefont {Y.}~\bibnamefont {{Ota}}}, \bibinfo {author}
  {\bibfnamefont {T.}~\bibnamefont {{Kondo}}}, \bibinfo {author} {\bibfnamefont
  {K.}~\bibnamefont {{Okazaki}}}, \bibinfo {author} {\bibfnamefont
  {Z.}~\bibnamefont {{Wang}}}, \bibinfo {author} {\bibfnamefont
  {J.}~\bibnamefont {{Wen}}}, \bibinfo {author} {\bibfnamefont {G.~D.}\
  \bibnamefont {{Gu}}}, \bibinfo {author} {\bibfnamefont {H.}~\bibnamefont
  {{Ding}}}, \ and\ \bibinfo {author} {\bibfnamefont {S.}~\bibnamefont
  {{Shin}}},\ }\href {\doibase 10.1126/science.aan4596} {\bibfield  {journal}
  {\bibinfo  {journal} {Science}\ }\textbf {\bibinfo {volume} {360}},\ \bibinfo
  {pages} {182} (\bibinfo {year} {2018})}\BibitemShut {NoStop}%
\bibitem [{\citenamefont {{Wang}}\ \emph {et~al.}(2018)\citenamefont {{Wang}},
  \citenamefont {{Kong}}, \citenamefont {{Fan}}, \citenamefont {{Chen}},
  \citenamefont {{Zhu}}, \citenamefont {{Liu}}, \citenamefont {{Cao}},
  \citenamefont {{Sun}}, \citenamefont {{Du}}, \citenamefont {{Schneeloch}},
  \citenamefont {{Zhong}}, \citenamefont {{Gu}}, \citenamefont {{Fu}},
  \citenamefont {{Ding}},\ and\ \citenamefont {{Gao}}}]{wang2018evidence}%
  \BibitemOpen
  \bibfield  {author} {\bibinfo {author} {\bibfnamefont {D.}~\bibnamefont
  {{Wang}}}, \bibinfo {author} {\bibfnamefont {L.}~\bibnamefont {{Kong}}},
  \bibinfo {author} {\bibfnamefont {P.}~\bibnamefont {{Fan}}}, \bibinfo
  {author} {\bibfnamefont {H.}~\bibnamefont {{Chen}}}, \bibinfo {author}
  {\bibfnamefont {S.}~\bibnamefont {{Zhu}}}, \bibinfo {author} {\bibfnamefont
  {W.}~\bibnamefont {{Liu}}}, \bibinfo {author} {\bibfnamefont
  {L.}~\bibnamefont {{Cao}}}, \bibinfo {author} {\bibfnamefont
  {Y.}~\bibnamefont {{Sun}}}, \bibinfo {author} {\bibfnamefont
  {S.}~\bibnamefont {{Du}}}, \bibinfo {author} {\bibfnamefont {J.}~\bibnamefont
  {{Schneeloch}}}, \bibinfo {author} {\bibfnamefont {R.}~\bibnamefont
  {{Zhong}}}, \bibinfo {author} {\bibfnamefont {G.}~\bibnamefont {{Gu}}},
  \bibinfo {author} {\bibfnamefont {L.}~\bibnamefont {{Fu}}}, \bibinfo {author}
  {\bibfnamefont {H.}~\bibnamefont {{Ding}}}, \ and\ \bibinfo {author}
  {\bibfnamefont {H.-J.}\ \bibnamefont {{Gao}}},\ }\href {\doibase
  10.1126/science.aao1797} {\bibfield  {journal} {\bibinfo  {journal}
  {Science}\ }\textbf {\bibinfo {volume} {362}},\ \bibinfo {pages} {333}
  (\bibinfo {year} {2018})}\BibitemShut {NoStop}%
\bibitem [{\citenamefont {{Zhang}}\ \emph {et~al.}(2019)\citenamefont
  {{Zhang}}, \citenamefont {{Wang}}, \citenamefont {{Wu}}, \citenamefont
  {{Yaji}}, \citenamefont {{Ishida}}, \citenamefont {{Kohama}}, \citenamefont
  {{Dai}}, \citenamefont {{Sun}}, \citenamefont {{Bareille}}, \citenamefont
  {{Kuroda}}, \citenamefont {{Kondo}}, \citenamefont {{Okazaki}}, \citenamefont
  {{Kindo}}, \citenamefont {{Wang}}, \citenamefont {{Jin}}, \citenamefont
  {{Hu}}, \citenamefont {{Thomale}}, \citenamefont {{Sumida}}, \citenamefont
  {{Wu}}, \citenamefont {{Miyamoto}}, \citenamefont {{Okuda}}, \citenamefont
  {{Ding}}, \citenamefont {{Gu}}, \citenamefont {{Tamegai}}, \citenamefont
  {{Kawakami}}, \citenamefont {{Sato}},\ and\ \citenamefont
  {{Shin}}}]{zhang2018multiple}%
  \BibitemOpen
  \bibfield  {author} {\bibinfo {author} {\bibfnamefont {P.}~\bibnamefont
  {{Zhang}}}, \bibinfo {author} {\bibfnamefont {Z.}~\bibnamefont {{Wang}}},
  \bibinfo {author} {\bibfnamefont {X.}~\bibnamefont {{Wu}}}, \bibinfo {author}
  {\bibfnamefont {K.}~\bibnamefont {{Yaji}}}, \bibinfo {author} {\bibfnamefont
  {Y.}~\bibnamefont {{Ishida}}}, \bibinfo {author} {\bibfnamefont
  {Y.}~\bibnamefont {{Kohama}}}, \bibinfo {author} {\bibfnamefont
  {G.}~\bibnamefont {{Dai}}}, \bibinfo {author} {\bibfnamefont
  {Y.}~\bibnamefont {{Sun}}}, \bibinfo {author} {\bibfnamefont
  {C.}~\bibnamefont {{Bareille}}}, \bibinfo {author} {\bibfnamefont
  {K.}~\bibnamefont {{Kuroda}}}, \bibinfo {author} {\bibfnamefont
  {T.}~\bibnamefont {{Kondo}}}, \bibinfo {author} {\bibfnamefont
  {K.}~\bibnamefont {{Okazaki}}}, \bibinfo {author} {\bibfnamefont
  {K.}~\bibnamefont {{Kindo}}}, \bibinfo {author} {\bibfnamefont
  {X.}~\bibnamefont {{Wang}}}, \bibinfo {author} {\bibfnamefont
  {C.}~\bibnamefont {{Jin}}}, \bibinfo {author} {\bibfnamefont
  {J.}~\bibnamefont {{Hu}}}, \bibinfo {author} {\bibfnamefont {R.}~\bibnamefont
  {{Thomale}}}, \bibinfo {author} {\bibfnamefont {K.}~\bibnamefont {{Sumida}}},
  \bibinfo {author} {\bibfnamefont {S.}~\bibnamefont {{Wu}}}, \bibinfo {author}
  {\bibfnamefont {K.}~\bibnamefont {{Miyamoto}}}, \bibinfo {author}
  {\bibfnamefont {T.}~\bibnamefont {{Okuda}}}, \bibinfo {author} {\bibfnamefont
  {H.}~\bibnamefont {{Ding}}}, \bibinfo {author} {\bibfnamefont {G.~D.}\
  \bibnamefont {{Gu}}}, \bibinfo {author} {\bibfnamefont {T.}~\bibnamefont
  {{Tamegai}}}, \bibinfo {author} {\bibfnamefont {T.}~\bibnamefont
  {{Kawakami}}}, \bibinfo {author} {\bibfnamefont {M.}~\bibnamefont {{Sato}}},
  \ and\ \bibinfo {author} {\bibfnamefont {S.}~\bibnamefont {{Shin}}},\ }\href
  {\doibase 10.1038/s41567-018-0280-z} {\bibfield  {journal} {\bibinfo
  {journal} {Nature Physics}\ }\textbf {\bibinfo {volume} {15}},\ \bibinfo
  {pages} {41} (\bibinfo {year} {2019})}\BibitemShut {NoStop}%
\bibitem [{\citenamefont {{Hao}}\ and\ \citenamefont
  {{Hu}}()}]{hao2018topological}%
  \BibitemOpen
  \bibfield  {author} {\bibinfo {author} {\bibfnamefont {N.}~\bibnamefont
  {{Hao}}}\ and\ \bibinfo {author} {\bibfnamefont {J.}~\bibnamefont {{Hu}}},\
  }\href@noop {} {\ }\Eprint {http://arxiv.org/abs/1811.03802}
  {arXiv:1811.03802 [cond-mat.mes-hall]} \BibitemShut {NoStop}%
\bibitem [{\citenamefont {{Machida}}\ \emph {et~al.}()\citenamefont
  {{Machida}}, \citenamefont {{Sun}}, \citenamefont {{Pyon}}, \citenamefont
  {{Takeda}}, \citenamefont {{Kohsaka}}, \citenamefont {{Hanaguri}},
  \citenamefont {{Sasagawa}},\ and\ \citenamefont
  {{Tamegai}}}]{machida2018zero}%
  \BibitemOpen
  \bibfield  {author} {\bibinfo {author} {\bibfnamefont {T.}~\bibnamefont
  {{Machida}}}, \bibinfo {author} {\bibfnamefont {Y.}~\bibnamefont {{Sun}}},
  \bibinfo {author} {\bibfnamefont {S.}~\bibnamefont {{Pyon}}}, \bibinfo
  {author} {\bibfnamefont {S.}~\bibnamefont {{Takeda}}}, \bibinfo {author}
  {\bibfnamefont {Y.}~\bibnamefont {{Kohsaka}}}, \bibinfo {author}
  {\bibfnamefont {T.}~\bibnamefont {{Hanaguri}}}, \bibinfo {author}
  {\bibfnamefont {T.}~\bibnamefont {{Sasagawa}}}, \ and\ \bibinfo {author}
  {\bibfnamefont {T.}~\bibnamefont {{Tamegai}}},\ }\href@noop {} {\ }\Eprint
  {http://arxiv.org/abs/1812.08995} {arXiv:1812.08995 [cond-mat.supr-con]}
  \BibitemShut {NoStop}%
\bibitem [{\citenamefont {Kong}\ \emph {et~al.}(2019)\citenamefont {Kong},
  \citenamefont {Zhu}, \citenamefont {Papaj}, \citenamefont {Cao},
  \citenamefont {Isobe}, \citenamefont {Liu}, \citenamefont {Wang},
  \citenamefont {Fan}, \citenamefont {Chen}, \citenamefont {Sun} \emph
  {et~al.}}]{kong2019observation}%
  \BibitemOpen
  \bibfield  {author} {\bibinfo {author} {\bibfnamefont {L.}~\bibnamefont
  {Kong}}, \bibinfo {author} {\bibfnamefont {S.}~\bibnamefont {Zhu}}, \bibinfo
  {author} {\bibfnamefont {M.}~\bibnamefont {Papaj}}, \bibinfo {author}
  {\bibfnamefont {L.}~\bibnamefont {Cao}}, \bibinfo {author} {\bibfnamefont
  {H.}~\bibnamefont {Isobe}}, \bibinfo {author} {\bibfnamefont
  {W.}~\bibnamefont {Liu}}, \bibinfo {author} {\bibfnamefont {D.}~\bibnamefont
  {Wang}}, \bibinfo {author} {\bibfnamefont {P.}~\bibnamefont {Fan}}, \bibinfo
  {author} {\bibfnamefont {H.}~\bibnamefont {Chen}}, \bibinfo {author}
  {\bibfnamefont {Y.}~\bibnamefont {Sun}},  \emph {et~al.},\ }\href@noop {}
  {\bibfield  {journal} {\bibinfo  {journal} {arXiv preprint arXiv:1901.02293}\
  } (\bibinfo {year} {2019})}\BibitemShut {NoStop}%
\bibitem [{\citenamefont {Zhu}\ \emph {et~al.}(2019)\citenamefont {Zhu},
  \citenamefont {Kong}, \citenamefont {Cao}, \citenamefont {Chen},
  \citenamefont {Du}, \citenamefont {Xing}, \citenamefont {Liu}, \citenamefont
  {Wang}, \citenamefont {Shen}, \citenamefont {Yang} \emph
  {et~al.}}]{zhu2019observation}%
  \BibitemOpen
  \bibfield  {author} {\bibinfo {author} {\bibfnamefont {S.}~\bibnamefont
  {Zhu}}, \bibinfo {author} {\bibfnamefont {L.}~\bibnamefont {Kong}}, \bibinfo
  {author} {\bibfnamefont {L.}~\bibnamefont {Cao}}, \bibinfo {author}
  {\bibfnamefont {H.}~\bibnamefont {Chen}}, \bibinfo {author} {\bibfnamefont
  {S.}~\bibnamefont {Du}}, \bibinfo {author} {\bibfnamefont {Y.}~\bibnamefont
  {Xing}}, \bibinfo {author} {\bibfnamefont {W.}~\bibnamefont {Liu}}, \bibinfo
  {author} {\bibfnamefont {D.}~\bibnamefont {Wang}}, \bibinfo {author}
  {\bibfnamefont {C.}~\bibnamefont {Shen}}, \bibinfo {author} {\bibfnamefont
  {F.}~\bibnamefont {Yang}},  \emph {et~al.},\ }\href@noop {} {\bibfield
  {journal} {\bibinfo  {journal} {arXiv preprint arXiv:1904.06124}\ } (\bibinfo
  {year} {2019})}\BibitemShut {NoStop}%
\bibitem [{\citenamefont {Fu}\ and\ \citenamefont
  {Kane}(2008)}]{fu2008superconducting}%
  \BibitemOpen
  \bibfield  {author} {\bibinfo {author} {\bibfnamefont {L.}~\bibnamefont
  {Fu}}\ and\ \bibinfo {author} {\bibfnamefont {C.~L.}\ \bibnamefont {Kane}},\
  }\href {\doibase 10.1103/PhysRevLett.100.096407} {\bibfield  {journal}
  {\bibinfo  {journal} {Phys. Rev. Lett.}\ }\textbf {\bibinfo {volume} {100}},\
  \bibinfo {pages} {096407} (\bibinfo {year} {2008})}\BibitemShut {NoStop}%
\bibitem [{\citenamefont {Shi}\ \emph {et~al.}(2017)\citenamefont {Shi},
  \citenamefont {Han}, \citenamefont {Richard}, \citenamefont {Wu},
  \citenamefont {Peng}, \citenamefont {Qian}, \citenamefont {Wang},
  \citenamefont {Hu}, \citenamefont {Sun},\ and\ \citenamefont
  {Ding}}]{shi2017fete1}%
  \BibitemOpen
  \bibfield  {author} {\bibinfo {author} {\bibfnamefont {X.}~\bibnamefont
  {Shi}}, \bibinfo {author} {\bibfnamefont {Z.-Q.}\ \bibnamefont {Han}},
  \bibinfo {author} {\bibfnamefont {P.}~\bibnamefont {Richard}}, \bibinfo
  {author} {\bibfnamefont {X.-X.}\ \bibnamefont {Wu}}, \bibinfo {author}
  {\bibfnamefont {X.-L.}\ \bibnamefont {Peng}}, \bibinfo {author}
  {\bibfnamefont {T.}~\bibnamefont {Qian}}, \bibinfo {author} {\bibfnamefont
  {S.-C.}\ \bibnamefont {Wang}}, \bibinfo {author} {\bibfnamefont {J.-P.}\
  \bibnamefont {Hu}}, \bibinfo {author} {\bibfnamefont {Y.-J.}\ \bibnamefont
  {Sun}}, \ and\ \bibinfo {author} {\bibfnamefont {H.}~\bibnamefont {Ding}},\
  }\href@noop {} {\bibfield  {journal} {\bibinfo  {journal} {Science bulletin}\
  }\textbf {\bibinfo {volume} {62}},\ \bibinfo {pages} {503} (\bibinfo {year}
  {2017})}\BibitemShut {NoStop}%
\bibitem [{\citenamefont {Peng}\ \emph {et~al.}(2019)\citenamefont {Peng},
  \citenamefont {Li}, \citenamefont {Wu}, \citenamefont {Deng}, \citenamefont
  {Shi}, \citenamefont {Fan}, \citenamefont {Li}, \citenamefont {Huang},
  \citenamefont {Qian}, \citenamefont {Richard} \emph
  {et~al.}}]{peng2019observation}%
  \BibitemOpen
  \bibfield  {author} {\bibinfo {author} {\bibfnamefont {X.-L.}\ \bibnamefont
  {Peng}}, \bibinfo {author} {\bibfnamefont {Y.}~\bibnamefont {Li}}, \bibinfo
  {author} {\bibfnamefont {X.-X.}\ \bibnamefont {Wu}}, \bibinfo {author}
  {\bibfnamefont {H.-B.}\ \bibnamefont {Deng}}, \bibinfo {author}
  {\bibfnamefont {X.}~\bibnamefont {Shi}}, \bibinfo {author} {\bibfnamefont
  {W.-H.}\ \bibnamefont {Fan}}, \bibinfo {author} {\bibfnamefont
  {M.}~\bibnamefont {Li}}, \bibinfo {author} {\bibfnamefont {Y.-B.}\
  \bibnamefont {Huang}}, \bibinfo {author} {\bibfnamefont {T.}~\bibnamefont
  {Qian}}, \bibinfo {author} {\bibfnamefont {P.}~\bibnamefont {Richard}},
  \emph {et~al.},\ }\href@noop {} {\bibfield  {journal} {\bibinfo  {journal}
  {arXiv preprint arXiv:1903.05968}\ } (\bibinfo {year} {2019})}\BibitemShut
  {NoStop}%
\bibitem [{\citenamefont {Qing-Yan}\ \emph {et~al.}(2012)\citenamefont
  {Qing-Yan}, \citenamefont {Zhi}, \citenamefont {Wen-Hao}, \citenamefont
  {Zuo-Cheng}, \citenamefont {Jin-Song}, \citenamefont {Wei}, \citenamefont
  {Hao}, \citenamefont {Yun-Bo}, \citenamefont {Peng}, \citenamefont {Kai}
  \emph {et~al.}}]{qing2012interface}%
  \BibitemOpen
  \bibfield  {author} {\bibinfo {author} {\bibfnamefont {W.}~\bibnamefont
  {Qing-Yan}}, \bibinfo {author} {\bibfnamefont {L.}~\bibnamefont {Zhi}},
  \bibinfo {author} {\bibfnamefont {Z.}~\bibnamefont {Wen-Hao}}, \bibinfo
  {author} {\bibfnamefont {Z.}~\bibnamefont {Zuo-Cheng}}, \bibinfo {author}
  {\bibfnamefont {Z.}~\bibnamefont {Jin-Song}}, \bibinfo {author}
  {\bibfnamefont {L.}~\bibnamefont {Wei}}, \bibinfo {author} {\bibfnamefont
  {D.}~\bibnamefont {Hao}}, \bibinfo {author} {\bibfnamefont {O.}~\bibnamefont
  {Yun-Bo}}, \bibinfo {author} {\bibfnamefont {D.}~\bibnamefont {Peng}},
  \bibinfo {author} {\bibfnamefont {C.}~\bibnamefont {Kai}},  \emph {et~al.},\
  }\href@noop {} {\bibfield  {journal} {\bibinfo  {journal} {Chinese Physics
  Letters}\ }\textbf {\bibinfo {volume} {29}},\ \bibinfo {pages} {037402}
  (\bibinfo {year} {2012})}\BibitemShut {NoStop}%
\bibitem [{\citenamefont {Wen-Hao}\ \emph {et~al.}(2014)\citenamefont
  {Wen-Hao}, \citenamefont {Yi}, \citenamefont {Jin-Song}, \citenamefont
  {Fang-Sen}, \citenamefont {Ming-Hua}, \citenamefont {Yan-Fei}, \citenamefont
  {Hui-Min}, \citenamefont {Jun-Ping}, \citenamefont {Ying}, \citenamefont
  {Hui-Chao} \emph {et~al.}}]{wen2014direct}%
  \BibitemOpen
  \bibfield  {author} {\bibinfo {author} {\bibfnamefont {Z.}~\bibnamefont
  {Wen-Hao}}, \bibinfo {author} {\bibfnamefont {S.}~\bibnamefont {Yi}},
  \bibinfo {author} {\bibfnamefont {Z.}~\bibnamefont {Jin-Song}}, \bibinfo
  {author} {\bibfnamefont {L.}~\bibnamefont {Fang-Sen}}, \bibinfo {author}
  {\bibfnamefont {G.}~\bibnamefont {Ming-Hua}}, \bibinfo {author}
  {\bibfnamefont {Z.}~\bibnamefont {Yan-Fei}}, \bibinfo {author} {\bibfnamefont
  {Z.}~\bibnamefont {Hui-Min}}, \bibinfo {author} {\bibfnamefont
  {P.}~\bibnamefont {Jun-Ping}}, \bibinfo {author} {\bibfnamefont
  {X.}~\bibnamefont {Ying}}, \bibinfo {author} {\bibfnamefont {W.}~\bibnamefont
  {Hui-Chao}},  \emph {et~al.},\ }\href@noop {} {\bibfield  {journal} {\bibinfo
   {journal} {Chinese Physics Letters}\ }\textbf {\bibinfo {volume} {31}},\
  \bibinfo {pages} {017401} (\bibinfo {year} {2014})}\BibitemShut {NoStop}%
\bibitem [{\citenamefont {Ge}\ \emph {et~al.}(2015)\citenamefont {Ge},
  \citenamefont {Liu}, \citenamefont {Liu}, \citenamefont {Gao}, \citenamefont
  {Qian}, \citenamefont {Xue}, \citenamefont {Liu},\ and\ \citenamefont
  {Jia}}]{ge2015superconductivity}%
  \BibitemOpen
  \bibfield  {author} {\bibinfo {author} {\bibfnamefont {J.-F.}\ \bibnamefont
  {Ge}}, \bibinfo {author} {\bibfnamefont {Z.-L.}\ \bibnamefont {Liu}},
  \bibinfo {author} {\bibfnamefont {C.}~\bibnamefont {Liu}}, \bibinfo {author}
  {\bibfnamefont {C.-L.}\ \bibnamefont {Gao}}, \bibinfo {author} {\bibfnamefont
  {D.}~\bibnamefont {Qian}}, \bibinfo {author} {\bibfnamefont {Q.-K.}\
  \bibnamefont {Xue}}, \bibinfo {author} {\bibfnamefont {Y.}~\bibnamefont
  {Liu}}, \ and\ \bibinfo {author} {\bibfnamefont {J.-F.}\ \bibnamefont
  {Jia}},\ }\href@noop {} {\bibfield  {journal} {\bibinfo  {journal} {Nature
  materials}\ }\textbf {\bibinfo {volume} {14}},\ \bibinfo {pages} {285}
  (\bibinfo {year} {2015})}\BibitemShut {NoStop}%
\bibitem [{\citenamefont {Qi}\ \emph {et~al.}(2009)\citenamefont {Qi},
  \citenamefont {Hughes}, \citenamefont {Raghu},\ and\ \citenamefont
  {Zhang}}]{qi2009time}%
  \BibitemOpen
  \bibfield  {author} {\bibinfo {author} {\bibfnamefont {X.-L.}\ \bibnamefont
  {Qi}}, \bibinfo {author} {\bibfnamefont {T.~L.}\ \bibnamefont {Hughes}},
  \bibinfo {author} {\bibfnamefont {S.}~\bibnamefont {Raghu}}, \ and\ \bibinfo
  {author} {\bibfnamefont {S.-C.}\ \bibnamefont {Zhang}},\ }\href {\doibase
  10.1103/PhysRevLett.102.187001} {\bibfield  {journal} {\bibinfo  {journal}
  {Phys. Rev. Lett.}\ }\textbf {\bibinfo {volume} {102}},\ \bibinfo {pages}
  {187001} (\bibinfo {year} {2009})}\BibitemShut {NoStop}%
\bibitem [{\citenamefont {Zhang}\ \emph
  {et~al.}(2013{\natexlab{a}})\citenamefont {Zhang}, \citenamefont {Kane},\
  and\ \citenamefont {Mele}}]{zhang2013time}%
  \BibitemOpen
  \bibfield  {author} {\bibinfo {author} {\bibfnamefont {F.}~\bibnamefont
  {Zhang}}, \bibinfo {author} {\bibfnamefont {C.~L.}\ \bibnamefont {Kane}}, \
  and\ \bibinfo {author} {\bibfnamefont {E.~J.}\ \bibnamefont {Mele}},\ }\href
  {\doibase 10.1103/PhysRevLett.111.056402} {\bibfield  {journal} {\bibinfo
  {journal} {Phys. Rev. Lett.}\ }\textbf {\bibinfo {volume} {111}},\ \bibinfo
  {pages} {056402} (\bibinfo {year} {2013}{\natexlab{a}})}\BibitemShut
  {NoStop}%
\bibitem [{\citenamefont {Huang}\ and\ \citenamefont
  {Hoffman}(2017)}]{huang2017monolayer}%
  \BibitemOpen
  \bibfield  {author} {\bibinfo {author} {\bibfnamefont {D.}~\bibnamefont
  {Huang}}\ and\ \bibinfo {author} {\bibfnamefont {J.~E.}\ \bibnamefont
  {Hoffman}},\ }\href@noop {} {\bibfield  {journal} {\bibinfo  {journal}
  {Annual Review of Condensed Matter Physics}\ }\textbf {\bibinfo {volume}
  {8}},\ \bibinfo {pages} {311} (\bibinfo {year} {2017})}\BibitemShut {NoStop}%
\bibitem [{\citenamefont {Ma}\ \emph {et~al.}(2009)\citenamefont {Ma},
  \citenamefont {Ji}, \citenamefont {Hu}, \citenamefont {Lu},\ and\
  \citenamefont {Xiang}}]{ma2009first}%
  \BibitemOpen
  \bibfield  {author} {\bibinfo {author} {\bibfnamefont {F.}~\bibnamefont
  {Ma}}, \bibinfo {author} {\bibfnamefont {W.}~\bibnamefont {Ji}}, \bibinfo
  {author} {\bibfnamefont {J.}~\bibnamefont {Hu}}, \bibinfo {author}
  {\bibfnamefont {Z.-Y.}\ \bibnamefont {Lu}}, \ and\ \bibinfo {author}
  {\bibfnamefont {T.}~\bibnamefont {Xiang}},\ }\href@noop {} {\bibfield
  {journal} {\bibinfo  {journal} {Physical review letters}\ }\textbf {\bibinfo
  {volume} {102}},\ \bibinfo {pages} {177003} (\bibinfo {year}
  {2009})}\BibitemShut {NoStop}%
\bibitem [{\citenamefont {Bao}\ \emph {et~al.}(2009)\citenamefont {Bao},
  \citenamefont {Qiu}, \citenamefont {Huang}, \citenamefont {Green},
  \citenamefont {Zajdel}, \citenamefont {Fitzsimmons}, \citenamefont
  {Zhernenkov}, \citenamefont {Chang}, \citenamefont {Fang}, \citenamefont
  {Qian} \emph {et~al.}}]{bao2009tunable}%
  \BibitemOpen
  \bibfield  {author} {\bibinfo {author} {\bibfnamefont {W.}~\bibnamefont
  {Bao}}, \bibinfo {author} {\bibfnamefont {Y.}~\bibnamefont {Qiu}}, \bibinfo
  {author} {\bibfnamefont {Q.}~\bibnamefont {Huang}}, \bibinfo {author}
  {\bibfnamefont {M.}~\bibnamefont {Green}}, \bibinfo {author} {\bibfnamefont
  {P.}~\bibnamefont {Zajdel}}, \bibinfo {author} {\bibfnamefont
  {M.}~\bibnamefont {Fitzsimmons}}, \bibinfo {author} {\bibfnamefont
  {M.}~\bibnamefont {Zhernenkov}}, \bibinfo {author} {\bibfnamefont
  {S.}~\bibnamefont {Chang}}, \bibinfo {author} {\bibfnamefont
  {M.}~\bibnamefont {Fang}}, \bibinfo {author} {\bibfnamefont {B.}~\bibnamefont
  {Qian}},  \emph {et~al.},\ }\href@noop {} {\bibfield  {journal} {\bibinfo
  {journal} {Physical review letters}\ }\textbf {\bibinfo {volume} {102}},\
  \bibinfo {pages} {247001} (\bibinfo {year} {2009})}\BibitemShut {NoStop}%
\bibitem [{\citenamefont {Li}\ \emph {et~al.}(2009)\citenamefont {Li},
  \citenamefont {de~La~Cruz}, \citenamefont {Huang}, \citenamefont {Chen},
  \citenamefont {Lynn}, \citenamefont {Hu}, \citenamefont {Huang},
  \citenamefont {Hsu}, \citenamefont {Yeh}, \citenamefont {Wu} \emph
  {et~al.}}]{li2009first}%
  \BibitemOpen
  \bibfield  {author} {\bibinfo {author} {\bibfnamefont {S.}~\bibnamefont
  {Li}}, \bibinfo {author} {\bibfnamefont {C.}~\bibnamefont {de~La~Cruz}},
  \bibinfo {author} {\bibfnamefont {Q.}~\bibnamefont {Huang}}, \bibinfo
  {author} {\bibfnamefont {Y.}~\bibnamefont {Chen}}, \bibinfo {author}
  {\bibfnamefont {J.}~\bibnamefont {Lynn}}, \bibinfo {author} {\bibfnamefont
  {J.}~\bibnamefont {Hu}}, \bibinfo {author} {\bibfnamefont {Y.-L.}\
  \bibnamefont {Huang}}, \bibinfo {author} {\bibfnamefont {F.-C.}\ \bibnamefont
  {Hsu}}, \bibinfo {author} {\bibfnamefont {K.-W.}\ \bibnamefont {Yeh}},
  \bibinfo {author} {\bibfnamefont {M.-K.}\ \bibnamefont {Wu}},  \emph
  {et~al.},\ }\href@noop {} {\bibfield  {journal} {\bibinfo  {journal}
  {Physical Review B}\ }\textbf {\bibinfo {volume} {79}},\ \bibinfo {pages}
  {054503} (\bibinfo {year} {2009})}\BibitemShut {NoStop}%
\bibitem [{\citenamefont {Manna}\ \emph {et~al.}(2017)\citenamefont {Manna},
  \citenamefont {Kamlapure}, \citenamefont {Cornils}, \citenamefont
  {H{\"a}nke}, \citenamefont {Hedegaard}, \citenamefont {Bremholm},
  \citenamefont {Iversen}, \citenamefont {Hofmann}, \citenamefont {Wiebe},\
  and\ \citenamefont {Wiesendanger}}]{manna2017interfacial}%
  \BibitemOpen
  \bibfield  {author} {\bibinfo {author} {\bibfnamefont {S.}~\bibnamefont
  {Manna}}, \bibinfo {author} {\bibfnamefont {A.}~\bibnamefont {Kamlapure}},
  \bibinfo {author} {\bibfnamefont {L.}~\bibnamefont {Cornils}}, \bibinfo
  {author} {\bibfnamefont {T.}~\bibnamefont {H{\"a}nke}}, \bibinfo {author}
  {\bibfnamefont {E.}~\bibnamefont {Hedegaard}}, \bibinfo {author}
  {\bibfnamefont {M.}~\bibnamefont {Bremholm}}, \bibinfo {author}
  {\bibfnamefont {B.}~\bibnamefont {Iversen}}, \bibinfo {author} {\bibfnamefont
  {P.}~\bibnamefont {Hofmann}}, \bibinfo {author} {\bibfnamefont
  {J.}~\bibnamefont {Wiebe}}, \ and\ \bibinfo {author} {\bibfnamefont
  {R.}~\bibnamefont {Wiesendanger}},\ }\href@noop {} {\bibfield  {journal}
  {\bibinfo  {journal} {Nature communications}\ }\textbf {\bibinfo {volume}
  {8}},\ \bibinfo {pages} {14074} (\bibinfo {year} {2017})}\BibitemShut
  {NoStop}%
\bibitem [{\citenamefont {Benalcazar}\ \emph
  {et~al.}(2017{\natexlab{a}})\citenamefont {Benalcazar}, \citenamefont
  {Bernevig},\ and\ \citenamefont {Hughes}}]{benalcazar2017quantized}%
  \BibitemOpen
  \bibfield  {author} {\bibinfo {author} {\bibfnamefont {W.~A.}\ \bibnamefont
  {Benalcazar}}, \bibinfo {author} {\bibfnamefont {B.~A.}\ \bibnamefont
  {Bernevig}}, \ and\ \bibinfo {author} {\bibfnamefont {T.~L.}\ \bibnamefont
  {Hughes}},\ }\href {\doibase 10.1126/science.aah6442} {\bibfield  {journal}
  {\bibinfo  {journal} {Science}\ }\textbf {\bibinfo {volume} {357}},\ \bibinfo
  {pages} {61} (\bibinfo {year} {2017}{\natexlab{a}})}\BibitemShut {NoStop}%
\bibitem [{\citenamefont {Benalcazar}\ \emph
  {et~al.}(2017{\natexlab{b}})\citenamefont {Benalcazar}, \citenamefont
  {Bernevig},\ and\ \citenamefont {Hughes}}]{benalcazar2017electric}%
  \BibitemOpen
  \bibfield  {author} {\bibinfo {author} {\bibfnamefont {W.~A.}\ \bibnamefont
  {Benalcazar}}, \bibinfo {author} {\bibfnamefont {B.~A.}\ \bibnamefont
  {Bernevig}}, \ and\ \bibinfo {author} {\bibfnamefont {T.~L.}\ \bibnamefont
  {Hughes}},\ }\href {\doibase 10.1103/PhysRevB.96.245115} {\bibfield
  {journal} {\bibinfo  {journal} {Phys. Rev. B}\ }\textbf {\bibinfo {volume}
  {96}},\ \bibinfo {pages} {245115} (\bibinfo {year}
  {2017}{\natexlab{b}})}\BibitemShut {NoStop}%
\bibitem [{\citenamefont {Zhang}\ \emph
  {et~al.}(2013{\natexlab{b}})\citenamefont {Zhang}, \citenamefont {Kane},\
  and\ \citenamefont {Mele}}]{zhang2013surface}%
  \BibitemOpen
  \bibfield  {author} {\bibinfo {author} {\bibfnamefont {F.}~\bibnamefont
  {Zhang}}, \bibinfo {author} {\bibfnamefont {C.~L.}\ \bibnamefont {Kane}}, \
  and\ \bibinfo {author} {\bibfnamefont {E.~J.}\ \bibnamefont {Mele}},\ }\href
  {\doibase 10.1103/PhysRevLett.110.046404} {\bibfield  {journal} {\bibinfo
  {journal} {Phys. Rev. Lett.}\ }\textbf {\bibinfo {volume} {110}},\ \bibinfo
  {pages} {046404} (\bibinfo {year} {2013}{\natexlab{b}})}\BibitemShut
  {NoStop}%
\bibitem [{\citenamefont {Schindler}\ \emph {et~al.}(2018)\citenamefont
  {Schindler}, \citenamefont {Cook}, \citenamefont {Vergniory}, \citenamefont
  {Wang}, \citenamefont {Parkin}, \citenamefont {Bernevig},\ and\ \citenamefont
  {Neupert}}]{schindler2018higher}%
  \BibitemOpen
  \bibfield  {author} {\bibinfo {author} {\bibfnamefont {F.}~\bibnamefont
  {Schindler}}, \bibinfo {author} {\bibfnamefont {A.~M.}\ \bibnamefont {Cook}},
  \bibinfo {author} {\bibfnamefont {M.~G.}\ \bibnamefont {Vergniory}}, \bibinfo
  {author} {\bibfnamefont {Z.}~\bibnamefont {Wang}}, \bibinfo {author}
  {\bibfnamefont {S.~S.}\ \bibnamefont {Parkin}}, \bibinfo {author}
  {\bibfnamefont {B.~A.}\ \bibnamefont {Bernevig}}, \ and\ \bibinfo {author}
  {\bibfnamefont {T.}~\bibnamefont {Neupert}},\ }\href {\doibase
  10.1126/sciadv.aat0346} {\bibfield  {journal} {\bibinfo  {journal} {Science
  advances}\ }\textbf {\bibinfo {volume} {4}},\ \bibinfo {pages} {eaat0346}
  (\bibinfo {year} {2018})}\BibitemShut {NoStop}%
\bibitem [{\citenamefont {Langbehn}\ \emph {et~al.}(2017)\citenamefont
  {Langbehn}, \citenamefont {Peng}, \citenamefont {Trifunovic}, \citenamefont
  {von Oppen},\ and\ \citenamefont {Brouwer}}]{langbehn2017reflection}%
  \BibitemOpen
  \bibfield  {author} {\bibinfo {author} {\bibfnamefont {J.}~\bibnamefont
  {Langbehn}}, \bibinfo {author} {\bibfnamefont {Y.}~\bibnamefont {Peng}},
  \bibinfo {author} {\bibfnamefont {L.}~\bibnamefont {Trifunovic}}, \bibinfo
  {author} {\bibfnamefont {F.}~\bibnamefont {von Oppen}}, \ and\ \bibinfo
  {author} {\bibfnamefont {P.~W.}\ \bibnamefont {Brouwer}},\ }\href {\doibase
  10.1103/PhysRevLett.119.246401} {\bibfield  {journal} {\bibinfo  {journal}
  {Phys. Rev. Lett.}\ }\textbf {\bibinfo {volume} {119}},\ \bibinfo {pages}
  {246401} (\bibinfo {year} {2017})}\BibitemShut {NoStop}%
\bibitem [{\citenamefont {Khalaf}(2018)}]{khalaf2018higher}%
  \BibitemOpen
  \bibfield  {author} {\bibinfo {author} {\bibfnamefont {E.}~\bibnamefont
  {Khalaf}},\ }\href {\doibase 10.1103/PhysRevB.97.205136} {\bibfield
  {journal} {\bibinfo  {journal} {Phys. Rev. B}\ }\textbf {\bibinfo {volume}
  {97}},\ \bibinfo {pages} {205136} (\bibinfo {year} {2018})}\BibitemShut
  {NoStop}%
\bibitem [{\citenamefont {Liu}\ \emph {et~al.}(2018)\citenamefont {Liu},
  \citenamefont {He},\ and\ \citenamefont {Nori}}]{liu2018majorana}%
  \BibitemOpen
  \bibfield  {author} {\bibinfo {author} {\bibfnamefont {T.}~\bibnamefont
  {Liu}}, \bibinfo {author} {\bibfnamefont {J.~J.}\ \bibnamefont {He}}, \ and\
  \bibinfo {author} {\bibfnamefont {F.}~\bibnamefont {Nori}},\ }\href {\doibase
  10.1103/PhysRevB.98.245413} {\bibfield  {journal} {\bibinfo  {journal} {Phys.
  Rev. B}\ }\textbf {\bibinfo {volume} {98}},\ \bibinfo {pages} {245413}
  (\bibinfo {year} {2018})}\BibitemShut {NoStop}%
\bibitem [{\citenamefont {{Peng}}\ and\ \citenamefont
  {{Xu}}()}]{peng2018proximity}%
  \BibitemOpen
  \bibfield  {author} {\bibinfo {author} {\bibfnamefont {Y.}~\bibnamefont
  {{Peng}}}\ and\ \bibinfo {author} {\bibfnamefont {Y.}~\bibnamefont {{Xu}}},\
  }\href@noop {} {\ }\Eprint {http://arxiv.org/abs/1809.09112}
  {arXiv:1809.09112 [cond-mat.mes-hall]} \BibitemShut {NoStop}%
\bibitem [{\citenamefont {Shapourian}\ \emph {et~al.}(2018)\citenamefont
  {Shapourian}, \citenamefont {Wang},\ and\ \citenamefont
  {Ryu}}]{shapourian2018topological}%
  \BibitemOpen
  \bibfield  {author} {\bibinfo {author} {\bibfnamefont {H.}~\bibnamefont
  {Shapourian}}, \bibinfo {author} {\bibfnamefont {Y.}~\bibnamefont {Wang}}, \
  and\ \bibinfo {author} {\bibfnamefont {S.}~\bibnamefont {Ryu}},\ }\href
  {\doibase 10.1103/PhysRevB.97.094508} {\bibfield  {journal} {\bibinfo
  {journal} {Phys. Rev. B}\ }\textbf {\bibinfo {volume} {97}},\ \bibinfo
  {pages} {094508} (\bibinfo {year} {2018})}\BibitemShut {NoStop}%
\bibitem [{\citenamefont {Volpez}\ \emph {et~al.}(2019)\citenamefont {Volpez},
  \citenamefont {Loss},\ and\ \citenamefont {Klinovaja}}]{volpez2019second}%
  \BibitemOpen
  \bibfield  {author} {\bibinfo {author} {\bibfnamefont {Y.}~\bibnamefont
  {Volpez}}, \bibinfo {author} {\bibfnamefont {D.}~\bibnamefont {Loss}}, \ and\
  \bibinfo {author} {\bibfnamefont {J.}~\bibnamefont {Klinovaja}},\ }\href@noop
  {} {\bibfield  {journal} {\bibinfo  {journal} {Physical Review Letters}\
  }\textbf {\bibinfo {volume} {122}},\ \bibinfo {pages} {126402} (\bibinfo
  {year} {2019})}\BibitemShut {NoStop}%
\bibitem [{\citenamefont {Wang}\ \emph
  {et~al.}(2018{\natexlab{a}})\citenamefont {Wang}, \citenamefont {Liu},
  \citenamefont {Lu},\ and\ \citenamefont {Zhang}}]{wang2018high}%
  \BibitemOpen
  \bibfield  {author} {\bibinfo {author} {\bibfnamefont {Q.}~\bibnamefont
  {Wang}}, \bibinfo {author} {\bibfnamefont {C.-C.}\ \bibnamefont {Liu}},
  \bibinfo {author} {\bibfnamefont {Y.-M.}\ \bibnamefont {Lu}}, \ and\ \bibinfo
  {author} {\bibfnamefont {F.}~\bibnamefont {Zhang}},\ }\href {\doibase
  10.1103/PhysRevLett.121.186801} {\bibfield  {journal} {\bibinfo  {journal}
  {Phys. Rev. Lett.}\ }\textbf {\bibinfo {volume} {121}},\ \bibinfo {pages}
  {186801} (\bibinfo {year} {2018}{\natexlab{a}})}\BibitemShut {NoStop}%
\bibitem [{\citenamefont {Yan}\ \emph {et~al.}(2018)\citenamefont {Yan},
  \citenamefont {Song},\ and\ \citenamefont {Wang}}]{yan2018majorana}%
  \BibitemOpen
  \bibfield  {author} {\bibinfo {author} {\bibfnamefont {Z.}~\bibnamefont
  {Yan}}, \bibinfo {author} {\bibfnamefont {F.}~\bibnamefont {Song}}, \ and\
  \bibinfo {author} {\bibfnamefont {Z.}~\bibnamefont {Wang}},\ }\href {\doibase
  10.1103/PhysRevLett.121.096803} {\bibfield  {journal} {\bibinfo  {journal}
  {Phys. Rev. Lett.}\ }\textbf {\bibinfo {volume} {121}},\ \bibinfo {pages}
  {096803} (\bibinfo {year} {2018})}\BibitemShut {NoStop}%
\bibitem [{\citenamefont {Wang}\ \emph
  {et~al.}(2018{\natexlab{b}})\citenamefont {Wang}, \citenamefont {Lin},\ and\
  \citenamefont {Hughes}}]{wang2018weak}%
  \BibitemOpen
  \bibfield  {author} {\bibinfo {author} {\bibfnamefont {Y.}~\bibnamefont
  {Wang}}, \bibinfo {author} {\bibfnamefont {M.}~\bibnamefont {Lin}}, \ and\
  \bibinfo {author} {\bibfnamefont {T.~L.}\ \bibnamefont {Hughes}},\ }\href
  {\doibase 10.1103/PhysRevB.98.165144} {\bibfield  {journal} {\bibinfo
  {journal} {Phys. Rev. B}\ }\textbf {\bibinfo {volume} {98}},\ \bibinfo
  {pages} {165144} (\bibinfo {year} {2018}{\natexlab{b}})}\BibitemShut
  {NoStop}%
\bibitem [{\citenamefont {Wu}\ \emph {et~al.}(2019{\natexlab{a}})\citenamefont
  {Wu}, \citenamefont {Yan},\ and\ \citenamefont {Huang}}]{wu2019higher}%
  \BibitemOpen
  \bibfield  {author} {\bibinfo {author} {\bibfnamefont {Z.}~\bibnamefont
  {Wu}}, \bibinfo {author} {\bibfnamefont {Z.}~\bibnamefont {Yan}}, \ and\
  \bibinfo {author} {\bibfnamefont {W.}~\bibnamefont {Huang}},\ }\href@noop {}
  {\bibfield  {journal} {\bibinfo  {journal} {Physical Review B}\ }\textbf
  {\bibinfo {volume} {99}},\ \bibinfo {pages} {020508} (\bibinfo {year}
  {2019}{\natexlab{a}})}\BibitemShut {NoStop}%
\bibitem [{\citenamefont {Zhu}(2018)}]{zhu2018tunable}%
  \BibitemOpen
  \bibfield  {author} {\bibinfo {author} {\bibfnamefont {X.}~\bibnamefont
  {Zhu}},\ }\href {\doibase 10.1103/PhysRevB.97.205134} {\bibfield  {journal}
  {\bibinfo  {journal} {Phys. Rev. B}\ }\textbf {\bibinfo {volume} {97}},\
  \bibinfo {pages} {205134} (\bibinfo {year} {2018})}\BibitemShut {NoStop}%
\bibitem [{\citenamefont {Bultinck}\ \emph {et~al.}(2019)\citenamefont
  {Bultinck}, \citenamefont {Bernevig},\ and\ \citenamefont
  {Zaletel}}]{bultinck2019three}%
  \BibitemOpen
  \bibfield  {author} {\bibinfo {author} {\bibfnamefont {N.}~\bibnamefont
  {Bultinck}}, \bibinfo {author} {\bibfnamefont {B.~A.}\ \bibnamefont
  {Bernevig}}, \ and\ \bibinfo {author} {\bibfnamefont {M.~P.}\ \bibnamefont
  {Zaletel}},\ }\href@noop {} {\bibfield  {journal} {\bibinfo  {journal}
  {Physical Review B}\ }\textbf {\bibinfo {volume} {99}},\ \bibinfo {pages}
  {125149} (\bibinfo {year} {2019})}\BibitemShut {NoStop}%
\bibitem [{\citenamefont {{Pan}}\ \emph {et~al.}()\citenamefont {{Pan}},
  \citenamefont {{Yang}}, \citenamefont {{Chen}}, \citenamefont {{Xu}},
  \citenamefont {{Liu}},\ and\ \citenamefont {{Liu}}}]{pan2018lattice}%
  \BibitemOpen
  \bibfield  {author} {\bibinfo {author} {\bibfnamefont {X.-H.}\ \bibnamefont
  {{Pan}}}, \bibinfo {author} {\bibfnamefont {K.-J.}\ \bibnamefont {{Yang}}},
  \bibinfo {author} {\bibfnamefont {L.}~\bibnamefont {{Chen}}}, \bibinfo
  {author} {\bibfnamefont {G.}~\bibnamefont {{Xu}}}, \bibinfo {author}
  {\bibfnamefont {C.-X.}\ \bibnamefont {{Liu}}}, \ and\ \bibinfo {author}
  {\bibfnamefont {X.}~\bibnamefont {{Liu}}},\ }\href@noop {} {\ }\Eprint
  {http://arxiv.org/abs/1812.10989} {arXiv:1812.10989 [cond-mat.mes-hall]}
  \BibitemShut {NoStop}%
\bibitem [{\citenamefont {Ghorashi}\ \emph {et~al.}(2019)\citenamefont
  {Ghorashi}, \citenamefont {Hu}, \citenamefont {Hughes},\ and\ \citenamefont
  {Rossi}}]{ghorashi2019second}%
  \BibitemOpen
  \bibfield  {author} {\bibinfo {author} {\bibfnamefont {S.~A.~A.}\
  \bibnamefont {Ghorashi}}, \bibinfo {author} {\bibfnamefont {X.}~\bibnamefont
  {Hu}}, \bibinfo {author} {\bibfnamefont {T.~L.}\ \bibnamefont {Hughes}}, \
  and\ \bibinfo {author} {\bibfnamefont {E.}~\bibnamefont {Rossi}},\
  }\href@noop {} {\bibfield  {journal} {\bibinfo  {journal} {arXiv preprint
  arXiv:1901.07579}\ } (\bibinfo {year} {2019})}\BibitemShut {NoStop}%
\bibitem [{\citenamefont {Franca}\ \emph {et~al.}(2019)\citenamefont {Franca},
  \citenamefont {Efremov},\ and\ \citenamefont {Fulga}}]{franca2019phase}%
  \BibitemOpen
  \bibfield  {author} {\bibinfo {author} {\bibfnamefont {S.}~\bibnamefont
  {Franca}}, \bibinfo {author} {\bibfnamefont {D.}~\bibnamefont {Efremov}}, \
  and\ \bibinfo {author} {\bibfnamefont {I.}~\bibnamefont {Fulga}},\
  }\href@noop {} {\bibfield  {journal} {\bibinfo  {journal} {arXiv preprint
  arXiv:1904.02437}\ } (\bibinfo {year} {2019})}\BibitemShut {NoStop}%
\bibitem [{\citenamefont {Hsu}\ \emph {et~al.}(2018)\citenamefont {Hsu},
  \citenamefont {Stano}, \citenamefont {Klinovaja},\ and\ \citenamefont
  {Loss}}]{hsu2018majorana}%
  \BibitemOpen
  \bibfield  {author} {\bibinfo {author} {\bibfnamefont {C.-H.}\ \bibnamefont
  {Hsu}}, \bibinfo {author} {\bibfnamefont {P.}~\bibnamefont {Stano}}, \bibinfo
  {author} {\bibfnamefont {J.}~\bibnamefont {Klinovaja}}, \ and\ \bibinfo
  {author} {\bibfnamefont {D.}~\bibnamefont {Loss}},\ }\href@noop {} {\bibfield
   {journal} {\bibinfo  {journal} {Physical review letters}\ }\textbf {\bibinfo
  {volume} {121}},\ \bibinfo {pages} {196801} (\bibinfo {year}
  {2018})}\BibitemShut {NoStop}%
\bibitem [{\citenamefont {Kudo}\ \emph {et~al.}(2019)\citenamefont {Kudo},
  \citenamefont {Yoshida},\ and\ \citenamefont {Hatsugai}}]{kudo2019higher}%
  \BibitemOpen
  \bibfield  {author} {\bibinfo {author} {\bibfnamefont {K.}~\bibnamefont
  {Kudo}}, \bibinfo {author} {\bibfnamefont {T.}~\bibnamefont {Yoshida}}, \
  and\ \bibinfo {author} {\bibfnamefont {Y.}~\bibnamefont {Hatsugai}},\
  }\href@noop {} {\bibfield  {journal} {\bibinfo  {journal} {arXiv preprint
  arXiv:1905.03484}\ } (\bibinfo {year} {2019})}\BibitemShut {NoStop}%
\bibitem [{\citenamefont {Yan}(2019)}]{yan2019higher}%
  \BibitemOpen
  \bibfield  {author} {\bibinfo {author} {\bibfnamefont {Z.}~\bibnamefont
  {Yan}},\ }\href@noop {} {\bibfield  {journal} {\bibinfo  {journal} {arXiv
  preprint arXiv:1905.11411}\ } (\bibinfo {year} {2019})}\BibitemShut {NoStop}%
\bibitem [{\citenamefont {Zhang}\ \emph {et~al.}(2019)\citenamefont {Zhang},
  \citenamefont {Cole},\ and\ \citenamefont {Sarma}}]{zhang2019helical}%
  \BibitemOpen
  \bibfield  {author} {\bibinfo {author} {\bibfnamefont {R.-X.}\ \bibnamefont
  {Zhang}}, \bibinfo {author} {\bibfnamefont {W.~S.}\ \bibnamefont {Cole}}, \
  and\ \bibinfo {author} {\bibfnamefont {S.~D.}\ \bibnamefont {Sarma}},\
  }\href@noop {} {\bibfield  {journal} {\bibinfo  {journal} {Physical Review
  Letters}\ }\textbf {\bibinfo {volume} {122}},\ \bibinfo {pages} {187001}
  (\bibinfo {year} {2019})}\BibitemShut {NoStop}%
\bibitem [{\citenamefont {Bernevig}\ \emph {et~al.}(2006)\citenamefont
  {Bernevig}, \citenamefont {Hughes},\ and\ \citenamefont
  {Zhang}}]{bernevig2006quantum}%
  \BibitemOpen
  \bibfield  {author} {\bibinfo {author} {\bibfnamefont {B.~A.}\ \bibnamefont
  {Bernevig}}, \bibinfo {author} {\bibfnamefont {T.~L.}\ \bibnamefont
  {Hughes}}, \ and\ \bibinfo {author} {\bibfnamefont {S.-C.}\ \bibnamefont
  {Zhang}},\ }\href@noop {} {\bibfield  {journal} {\bibinfo  {journal}
  {Science}\ }\textbf {\bibinfo {volume} {314}},\ \bibinfo {pages} {1757}
  (\bibinfo {year} {2006})}\BibitemShut {NoStop}%
\bibitem [{sup()}]{supplementary}%
  \BibitemOpen
  \href@noop {} {}\bibinfo {note} {See Supplemental Material at XX for the full
  Hamiltonian $H_\text{FTS}$, a derivation of topological condition, definition
  of topological charge for nodal points, an estimate of interlayer exchange
  coupling from DFT calculation, and a detailed analysis of magnetic disorder
  effects}\BibitemShut {NoStop}%
\bibitem [{\citenamefont {Mong}\ \emph {et~al.}(2010)\citenamefont {Mong},
  \citenamefont {Essin},\ and\ \citenamefont
  {Moore}}]{mong2010antiferromagnetic}%
  \BibitemOpen
  \bibfield  {author} {\bibinfo {author} {\bibfnamefont {R.~S.}\ \bibnamefont
  {Mong}}, \bibinfo {author} {\bibfnamefont {A.~M.}\ \bibnamefont {Essin}}, \
  and\ \bibinfo {author} {\bibfnamefont {J.~E.}\ \bibnamefont {Moore}},\
  }\href@noop {} {\bibfield  {journal} {\bibinfo  {journal} {Physical Review
  B}\ }\textbf {\bibinfo {volume} {81}},\ \bibinfo {pages} {245209} (\bibinfo
  {year} {2010})}\BibitemShut {NoStop}%
\bibitem [{\citenamefont {Liu}(2013)}]{liu2013antiferromagnetic}%
  \BibitemOpen
  \bibfield  {author} {\bibinfo {author} {\bibfnamefont {C.-X.}\ \bibnamefont
  {Liu}},\ }\href@noop {} {\bibfield  {journal} {\bibinfo  {journal} {arXiv
  preprint arXiv:1304.6455}\ } (\bibinfo {year} {2013})}\BibitemShut {NoStop}%
\bibitem [{\citenamefont {Zhang}\ and\ \citenamefont
  {Liu}(2015)}]{zhang2015topological}%
  \BibitemOpen
  \bibfield  {author} {\bibinfo {author} {\bibfnamefont {R.-X.}\ \bibnamefont
  {Zhang}}\ and\ \bibinfo {author} {\bibfnamefont {C.-X.}\ \bibnamefont
  {Liu}},\ }\href@noop {} {\bibfield  {journal} {\bibinfo  {journal} {Physical
  Review B}\ }\textbf {\bibinfo {volume} {91}},\ \bibinfo {pages} {115317}
  (\bibinfo {year} {2015})}\BibitemShut {NoStop}%
\bibitem [{\citenamefont {Fu}\ and\ \citenamefont
  {Kane}(2009)}]{fu2009josephson}%
  \BibitemOpen
  \bibfield  {author} {\bibinfo {author} {\bibfnamefont {L.}~\bibnamefont
  {Fu}}\ and\ \bibinfo {author} {\bibfnamefont {C.~L.}\ \bibnamefont {Kane}},\
  }\href@noop {} {\bibfield  {journal} {\bibinfo  {journal} {Physical Review
  B}\ }\textbf {\bibinfo {volume} {79}},\ \bibinfo {pages} {161408} (\bibinfo
  {year} {2009})}\BibitemShut {NoStop}%
\bibitem [{\citenamefont {{Alicea}}(2012)}]{alicea2012new}%
  \BibitemOpen
  \bibfield  {author} {\bibinfo {author} {\bibfnamefont {J.}~\bibnamefont
  {{Alicea}}},\ }\href {\doibase 10.1088/0034-4885/75/7/076501} {\bibfield
  {journal} {\bibinfo  {journal} {Reports on Progress in Physics}\ }\textbf
  {\bibinfo {volume} {75}},\ \bibinfo {eid} {076501} (\bibinfo {year}
  {2012})}\BibitemShut {NoStop}%
\bibitem [{\citenamefont {Schnyder}\ and\ \citenamefont
  {Ryu}(2011)}]{schnyder2011topological}%
  \BibitemOpen
  \bibfield  {author} {\bibinfo {author} {\bibfnamefont {A.~P.}\ \bibnamefont
  {Schnyder}}\ and\ \bibinfo {author} {\bibfnamefont {S.}~\bibnamefont {Ryu}},\
  }\href@noop {} {\bibfield  {journal} {\bibinfo  {journal} {Physical Review
  B}\ }\textbf {\bibinfo {volume} {84}},\ \bibinfo {pages} {060504} (\bibinfo
  {year} {2011})}\BibitemShut {NoStop}%
\bibitem [{\citenamefont {Yu}\ and\ \citenamefont {Liu}(2018)}]{yu2018singlet}%
  \BibitemOpen
  \bibfield  {author} {\bibinfo {author} {\bibfnamefont {J.}~\bibnamefont
  {Yu}}\ and\ \bibinfo {author} {\bibfnamefont {C.-X.}\ \bibnamefont {Liu}},\
  }\href@noop {} {\bibfield  {journal} {\bibinfo  {journal} {Physical Review
  B}\ }\textbf {\bibinfo {volume} {98}},\ \bibinfo {pages} {104514} (\bibinfo
  {year} {2018})}\BibitemShut {NoStop}%
\bibitem [{\citenamefont {Sun}\ \emph {et~al.}(2014)\citenamefont {Sun},
  \citenamefont {Zhang}, \citenamefont {Xing}, \citenamefont {Li},
  \citenamefont {Zhao}, \citenamefont {Xia}, \citenamefont {Wang},
  \citenamefont {Ma}, \citenamefont {Xue},\ and\ \citenamefont
  {Wang}}]{sun2014high}%
  \BibitemOpen
  \bibfield  {author} {\bibinfo {author} {\bibfnamefont {Y.}~\bibnamefont
  {Sun}}, \bibinfo {author} {\bibfnamefont {W.}~\bibnamefont {Zhang}}, \bibinfo
  {author} {\bibfnamefont {Y.}~\bibnamefont {Xing}}, \bibinfo {author}
  {\bibfnamefont {F.}~\bibnamefont {Li}}, \bibinfo {author} {\bibfnamefont
  {Y.}~\bibnamefont {Zhao}}, \bibinfo {author} {\bibfnamefont {Z.}~\bibnamefont
  {Xia}}, \bibinfo {author} {\bibfnamefont {L.}~\bibnamefont {Wang}}, \bibinfo
  {author} {\bibfnamefont {X.}~\bibnamefont {Ma}}, \bibinfo {author}
  {\bibfnamefont {Q.-K.}\ \bibnamefont {Xue}}, \ and\ \bibinfo {author}
  {\bibfnamefont {J.}~\bibnamefont {Wang}},\ }\href@noop {} {\bibfield
  {journal} {\bibinfo  {journal} {Scientific reports}\ }\textbf {\bibinfo
  {volume} {4}},\ \bibinfo {pages} {6040} (\bibinfo {year} {2014})}\BibitemShut
  {NoStop}%
\bibitem [{\citenamefont {Nabeshima}\ \emph {et~al.}(2017)\citenamefont
  {Nabeshima}, \citenamefont {Imai}, \citenamefont {Ichinose}, \citenamefont
  {Tsukada},\ and\ \citenamefont {Maeda}}]{nabeshima2017growth}%
  \BibitemOpen
  \bibfield  {author} {\bibinfo {author} {\bibfnamefont {F.}~\bibnamefont
  {Nabeshima}}, \bibinfo {author} {\bibfnamefont {Y.}~\bibnamefont {Imai}},
  \bibinfo {author} {\bibfnamefont {A.}~\bibnamefont {Ichinose}}, \bibinfo
  {author} {\bibfnamefont {I.}~\bibnamefont {Tsukada}}, \ and\ \bibinfo
  {author} {\bibfnamefont {A.}~\bibnamefont {Maeda}},\ }\href@noop {}
  {\bibfield  {journal} {\bibinfo  {journal} {Japanese Journal of Applied
  Physics}\ }\textbf {\bibinfo {volume} {56}},\ \bibinfo {pages} {020308}
  (\bibinfo {year} {2017})}\BibitemShut {NoStop}%
\bibitem [{\citenamefont {Zhang}\ \emph {et~al.}(2016)\citenamefont {Zhang},
  \citenamefont {Lee}, \citenamefont {Moore}, \citenamefont {Li}, \citenamefont
  {Yi}, \citenamefont {Hashimoto}, \citenamefont {Lu}, \citenamefont
  {Devereaux}, \citenamefont {Lee},\ and\ \citenamefont
  {Shen}}]{zhang2016superconducting}%
  \BibitemOpen
  \bibfield  {author} {\bibinfo {author} {\bibfnamefont {Y.}~\bibnamefont
  {Zhang}}, \bibinfo {author} {\bibfnamefont {J.}~\bibnamefont {Lee}}, \bibinfo
  {author} {\bibfnamefont {R.}~\bibnamefont {Moore}}, \bibinfo {author}
  {\bibfnamefont {W.}~\bibnamefont {Li}}, \bibinfo {author} {\bibfnamefont
  {M.}~\bibnamefont {Yi}}, \bibinfo {author} {\bibfnamefont {M.}~\bibnamefont
  {Hashimoto}}, \bibinfo {author} {\bibfnamefont {D.}~\bibnamefont {Lu}},
  \bibinfo {author} {\bibfnamefont {T.}~\bibnamefont {Devereaux}}, \bibinfo
  {author} {\bibfnamefont {D.-H.}\ \bibnamefont {Lee}}, \ and\ \bibinfo
  {author} {\bibfnamefont {Z.-X.}\ \bibnamefont {Shen}},\ }\href@noop {}
  {\bibfield  {journal} {\bibinfo  {journal} {Physical review letters}\
  }\textbf {\bibinfo {volume} {117}},\ \bibinfo {pages} {117001} (\bibinfo
  {year} {2016})}\BibitemShut {NoStop}%
\bibitem [{\citenamefont {Wu}\ \emph {et~al.}(2019{\natexlab{b}})\citenamefont
  {Wu}, \citenamefont {Liu}, \citenamefont {Thomale},\ and\ \citenamefont
  {Liu}}]{wu2019high}%
  \BibitemOpen
  \bibfield  {author} {\bibinfo {author} {\bibfnamefont {X.}~\bibnamefont
  {Wu}}, \bibinfo {author} {\bibfnamefont {X.}~\bibnamefont {Liu}}, \bibinfo
  {author} {\bibfnamefont {R.}~\bibnamefont {Thomale}}, \ and\ \bibinfo
  {author} {\bibfnamefont {C.-X.}\ \bibnamefont {Liu}},\ }\href@noop {}
  {\bibfield  {journal} {\bibinfo  {journal} {arXiv preprint arXiv:1905.10648}\
  } (\bibinfo {year} {2019}{\natexlab{b}})}\BibitemShut {NoStop}%
\end{thebibliography}%

\onecolumngrid

\subsection{\large Supplemental Material for ``Higher Order Topology and Nodal Topological Superconductivity in Fe(Se,Te) Heterostructures"}

\section{Full Hamiltonian with Superconductivity and Bicollinear Antiferromagnetism}

The full effective model for the FTS/FT heterostructure is
	\bea
	H_{\rm FTS} &=& {\cal G}_1 \otimes [\frac{A}{2I}(e^{ik_x}\Gamma_{167}-e^{ik_y}\Gamma_{2})+B(e^{ik_x}+e^{ik_y})\Gamma_5]  - {\cal G}_2 \otimes [\frac{A}{2I}(e^{-ik_x}\Gamma_{167}-e^{-ik_y}\Gamma_2)-B(e^{-ik_x}+e^{-ik_y})\Gamma_5] \nonumber \\
	&&+ {\cal G}_0 \otimes [(m-4B) \Gamma_5 +  \Delta \Gamma_{137} -\mu  \Gamma_{67}] + \frac{M}{\sqrt{1+\alpha^2}} {\cal G}_3 \otimes \tau_z \otimes  (\sigma_0 + \alpha \sigma_z) \otimes s_x.
	\label{Eq: FTS Hamiltonian}
	\eea
The ${\cal G}$ matrices describe the sublattice degree of freedom; the hopping matrix elements are described by
\bea
{\cal G}_1 = \begin{pmatrix}
	0 & 1 & 0 & 0 \\
	0 & 0 & 1 & 0 \\
	0 & 0 & 0 & 1 \\
	1 & 0 & 0 & 0 \\
\end{pmatrix},\ 
{\cal G}_2 = \begin{pmatrix}
	0 & 0 & 0 & 1 \\
	1 & 0 & 0 & 0 \\
	0 & 1 & 0 & 0 \\
	0 & 0 & 1 & 0 \\
\end{pmatrix},
\eea 
and the onsite matrix elements are described by diagonal matrices ${\cal G}_0 = \text{diag}[1,1,1,1]$ and ${\cal G}_3=\text{diag}[1,1,-1,-1]$. Here $M$ deontes the interlayer exchange coupling between FTS and FT. In particular, ${\cal G}_3$ describes opposite coupling for electrons with sublattice $i=1,2$ and $i=3,4$, which captures the bicollinear AFM texture.

For convenience, we have assumed the alignment of magnetic moments of FeTe to be along the $\pm {\bf x}$ direction. The model parameter $\alpha \in [-1,1]$ accounts for the possible difference in $g$ factors of $p$-electrons and $d$-electrons. Numerically, we find that changing $\alpha$ gives little contribution to the physics we are interested in, so without loss of generality we take $\alpha=0$ in our discussion.

\section{Topological Criterion for Higher Order Topology}

The key to realize corner Majorana physics is to understand the competition between FM and SC on the $\tilde{\bf y}$ edge. It is instuctive to start from an effective model of the edge theory
\bea
H_{\tilde{\bf y}} = k_{\tilde{y}} \tau_0 \otimes s_z + \delta_M \tau_z \otimes s_x + \Delta \tau_y\otimes s_y - \mu \tau_z \otimes s_0.
\label{Eq: Edge Theory}
\eea
Here $\delta_M$ is the induced magnetic gap on $\tilde{\bf y}$ edge, which is generally smaller than the exchange coupling effect $M$ in the bulk. For a given $M$, the value of $\delta_M$ can be identified by calculating the energy gap on the $\tilde{\bf y}$ edge at $\mu=\Delta=0$. In fact, we have numerically confirmed the existence of a simple linear relation between $M$ and $\delta_M$,
\bea
\delta_M \approx \beta_M M.
\eea
For our choice of parameters with $A=B=1,m=2$, we find that 
\bea
\beta_M = 0.678...
\eea
The eigenvalues of $H_{\tilde{\bf y}}$ can be solved analytically,
\bea
E=\pm\sqrt{k_{\tilde{y}}^2+\delta_M^2+\Delta^2+\mu^2 \pm 2\sqrt{m^2(\Delta^2+\mu^2)+k^2\mu^2}}.
\eea
Therefore, the edge topological phase transition happens when the edge energy gap closes. This is equivalent to finding $k_{\tilde{y}}$ such that $E=0$ is satisfied. It is straightforward to show that the equation we are solving is
\bea
k_{\tilde{y}}^4 + 2 k_{\tilde{y}}^2 (\delta_M^2+\Delta^2-\mu^2) + (\delta_M^2-\Delta^2-\mu^2)^2 =0,
\eea
which leads to 
\bea
k_{\tilde{y}}^2 &=& - [(\delta_M^2 -\mu^2) + \Delta^2 \pm 2\sqrt{\Delta^2 (\delta_M^2 - \mu^2)}] \nonumber \\
&=& -(\sqrt{\delta_M^2-\mu^2}\pm |\Delta|)^2 \nonumber \\
&\geq& 0.
\eea
Therefore, the topological phase transition can only happen at $k_{\tilde{y}}=0$ when
\bea
\delta_M^2 &=& \mu^2 + \Delta^2.
\eea
This leads to the topological criterion shown in the main text
\bea
M^2 > \frac{1}{\beta_M^2} (\mu^2 + \Delta^2).
\eea

\section{Topological Charge for Topological Nodal Superconductivity}

\begin{figure}[t]
	\centering
	\includegraphics[width=0.95\textwidth]{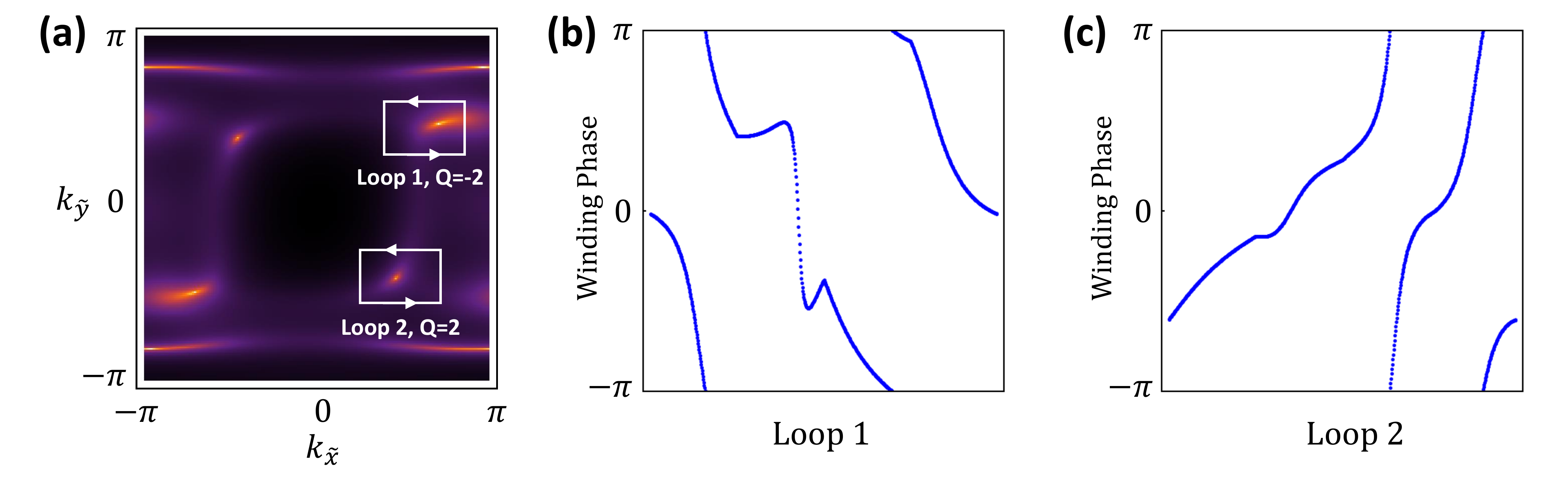}
	\caption{(a) Two closed loops that encloses nodal points are shown in white lines. The directions of the loops are shown by white arrows. (b) and (c) show the evolution of the winding phase for loop 1 and loop 2, respectively. Clearly, the nodal point enclosed by loop 1 (loop 2) has a topological charge of $Q=-2$ ($Q=+2$).}
	\label{Fig: Topological Charge}
\end{figure}

As discussed in the main text, the effective TRS symmetry $\Theta_M$ and particle-hole symmetry $\Pi$ lead to an emergent AFM chiral symmetry ${\cal C}=\Theta_M\Pi$. In the nodal SC phase, a topological charge $Q\in\mathbb{Z}$ can be defined based on ${\cal C}$ to characterize the topological nature of the nodal point. Given $\{{\cal C},H_\text{FTS}\}=0$, we notice that the unitary transformation $U_{\cal C}$ that diagonalizes ${\cal C}$ makes $H_\text{FTS}$ off-block-diagonal,
\bea
U_{\cal C} H_\text{FTS} U_{\cal C}^{\dagger} = \begin{pmatrix}
	0 & N(k) \\
	N(k)^{\dagger} & 0 \\
\end{pmatrix}.
\eea
We then perform a singular value decomposition to $N(k)$,
\bea
N(k) = {\cal U} (k) \Sigma (k) {\cal V}^{\dagger} (k)
\eea
and define 
\bea
{\cal D}(k) = {\cal U}(k) {\cal V}^{\dagger}(k).
\eea
The topological charge is simply the winding number of $\det N(k)$ along a closed loop ${\cal L}$ that encloses the nodal point, which is mathematically \cite{schnyder2011topological,yu2018singlet}
\bea
Q = \frac{1}{2\pi} \oint_{\cal L} d{\bf k} \cdot \nabla_k \text{Arg}[\det {\cal D}(k)].
\label{Eq: Winding Phase}
\eea
As shown in Fig. \ref{Fig: Topological Charge} (a), we have studied the topological charge of the nodal points for the same nodal phase in Fig. 3 of the main text. We have chosen two counter-clockwise closed loops (white lines) that encloses two inequivalent nodal points to study their winding number. To better visualize the winding number, we define a 1d momenta $k_{\cal L}$ along the loop ${\cal L}$ to parametrize the loop, and further define a winding phase at each $k_{\cal L}$, 
\bea
{\cal A}(k_{\cal L}) = \partial{k_{\cal L}} \text{Arg}[\det {\cal D}(k_{\cal L})]
\eea
to track the evolution of winding phase integral in Eq. \ref{Eq: Winding Phase}. As shown in Fig. \ref{Fig: Topological Charge} (b) and (c), the evolution of the winding phase for loop 1 and loop 2 clearly show the topological charge $Q$ of the enclosed nodal points to be $-2$ and $+2$, respectively.

\section{Estimate on the Proximity Induced Exchange Coupling in Fe(Te,Se) Monolayers}

We perform first principles calculations to obtain an estimate on the proximity induced exchange coupling in FeTe/Fe(Te,Se) heterostructures. For our purpose, we calculate the energy spectrum of a bilayer FeSe with an experimental lattice constant $a=3.905$~\AA~. In particular, the FeSe in the bottom layer is assumed to be ferromagnetic, while the top layer FeSe is initially non-magnetic. Consequently, the spin splitting for the top layer FeSe in the energy spectrum is expected to provide an approximate strength of the proximity induced exchange coupling effect.

\begin{figure}[H]
	\centering
	\includegraphics[width=0.5\textwidth]{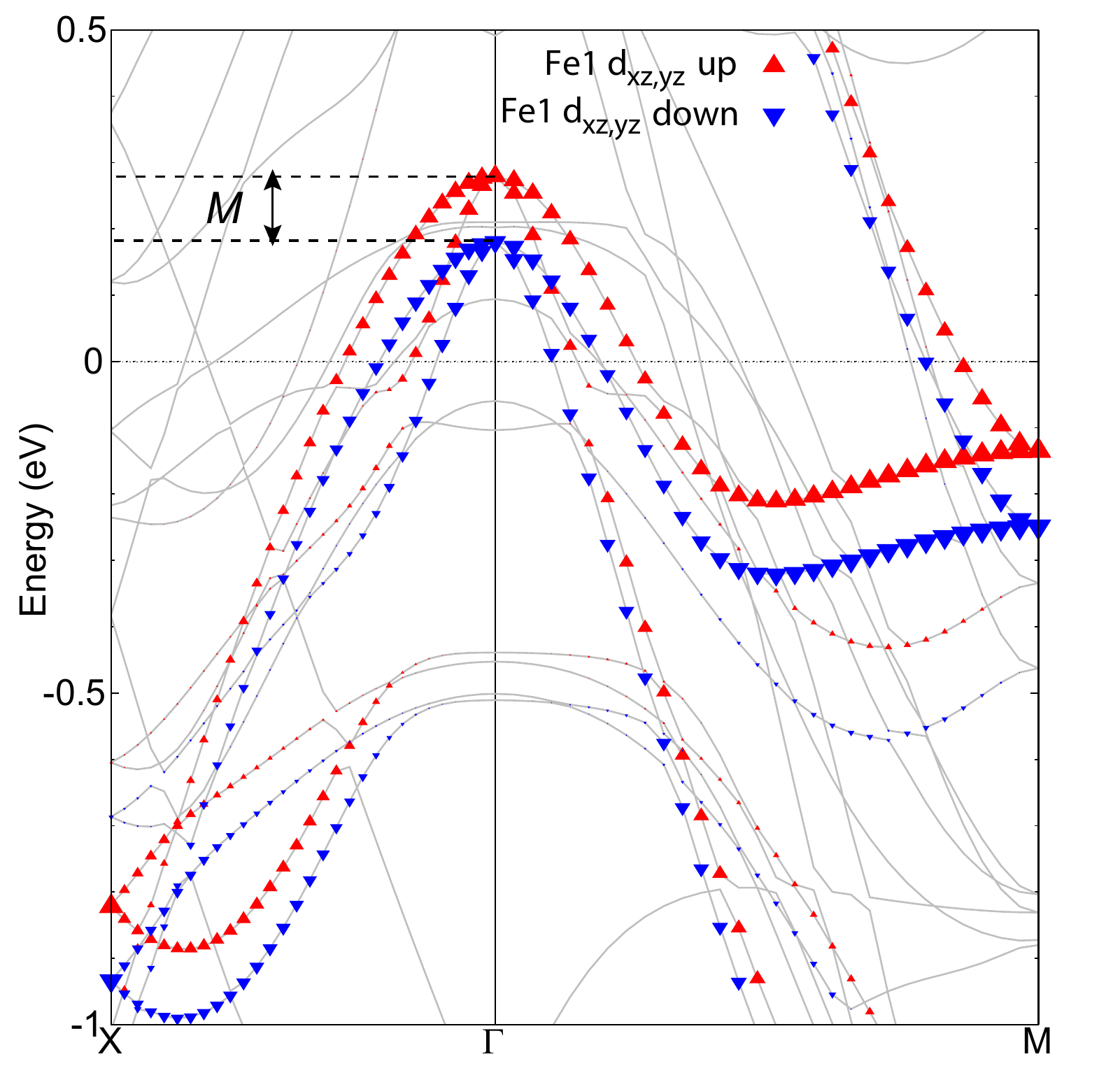}
	\caption{Band structure for the bilayer FeSe. Initially the bottom layer is ferromagnetic and the top layer is nonmagneitc. The up and down triangles denotes the $d_{xz}/d_{yz}$ states for the top FeSe layer. The induced spin splitting is found to be about 100 meV
		\label{band}}
\end{figure}

The obtained band structure is displayed in Fig. \ref{band}, where the up and down triangles denote the spin up and spin down $d_{xz}/d_{yz}$ states of the top FeSe layer. We find that the magnetic moment of the bottom FeSe layer is about 2.75 $\mu_B$ and the obtained exchange splitting for the top FeSe layer is about 100 meV (as shown by the spacing between black dashed lines in Fig. \ref{band}). Considering the known overestimation of magnetic moment for Fe in first principle calculations, we use the measured magnetic moment in neutron scattering experiments (1.65 $\mu_B$ along b axis, parallel spin axis)\cite{bao2009tunable} to further rescale our first principles results, which leads to an induced exchange coupling of 60 meV for the top FeSe layer.

\section{Stability of Higher Order Topology against Magnetic Disorder}

An important disorder mechanism for our proposed Fe(Se,Te) heterostructure arises from individual flipped spins (relative to the perfectly ordered ground state) in the antiferromagnetic layer. Although previous work has established some robustness of various higher-order topological phases against weak chemical potential disorder, the energy scale associated with a flipped spin in the AFM is necessarily the exchange coupling $M$, which we expect to be relatively large. Therefore, one can reasonably question if the HOTSC phase we predict requires perfect ordering of the bicollinear AFM. To address this issue, we numerically diagonalize $H_{\rm FTS}$ in real space for magnetic textures with a fixed fraction $n_{\rm imp}$ of ``impurity" sites, i.e., randomly selected, uncorrelated lattice points where the coupling $\pm M \rightarrow \mp M$. Following the main text, this is done for a $20 \tilde{a}_y \times 10 \tilde{a}_x$ system, with $(M, \Delta, \mu) = (0.6, 0.2, 0.0)$.

\begin{figure}[t]
	\centering
	\includegraphics[width=0.4\textwidth]{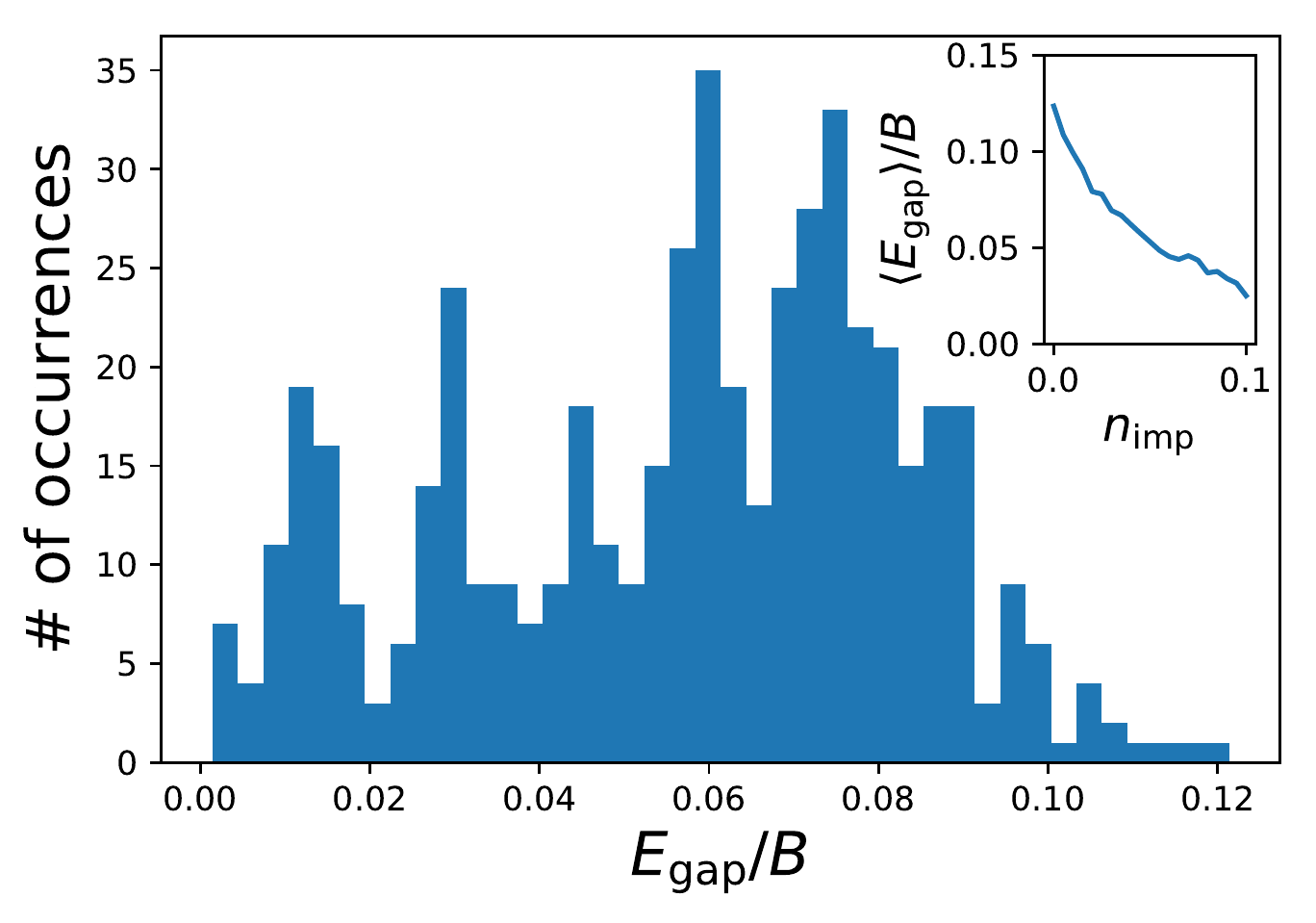}
	\caption{Distribution of gaps for 500 disorder realizations at $n_{\rm imp}=0.05$. The distribution is irregular, but demonstrates that additional quasiparticle states at all energies including zero can be generated by disorder. The inset shows the expected downward trend but nonzero value of the disorder averaged gap.
		\label{fig:SI_dis1}}
\end{figure}

We first characterize the effect of disorder on the edge gap in the spectrum, defined here as the energy of the lowest quasiparticle state above the corner states. In the inset of Fig.~\ref{fig:SI_dis1} we plot the \textit{disorder averaged} edge gap as a function of $n_{\rm imp}$. This average systematically trends downward as one might have expected a priori. More interesting, though, is that in the main panel we histogram the distribution of edge gaps over 500 independent disorder realizations at $n_{\rm imp}=0.05$, and see that it is generally irregular with a few distinct peaks. In other words, most disorder realizations bind low-energy (including zero energy) states well below the clean edge gap $\sim 0.12 B$. Although the \textit{average} gap is necessarily nonzero, the interplay of these individual low-energy states with the Majoranas must be investigated further to see if particular disorder realizations can trivialize them.

\begin{figure}[t]
	\centering
	\includegraphics[width=0.4\textwidth]{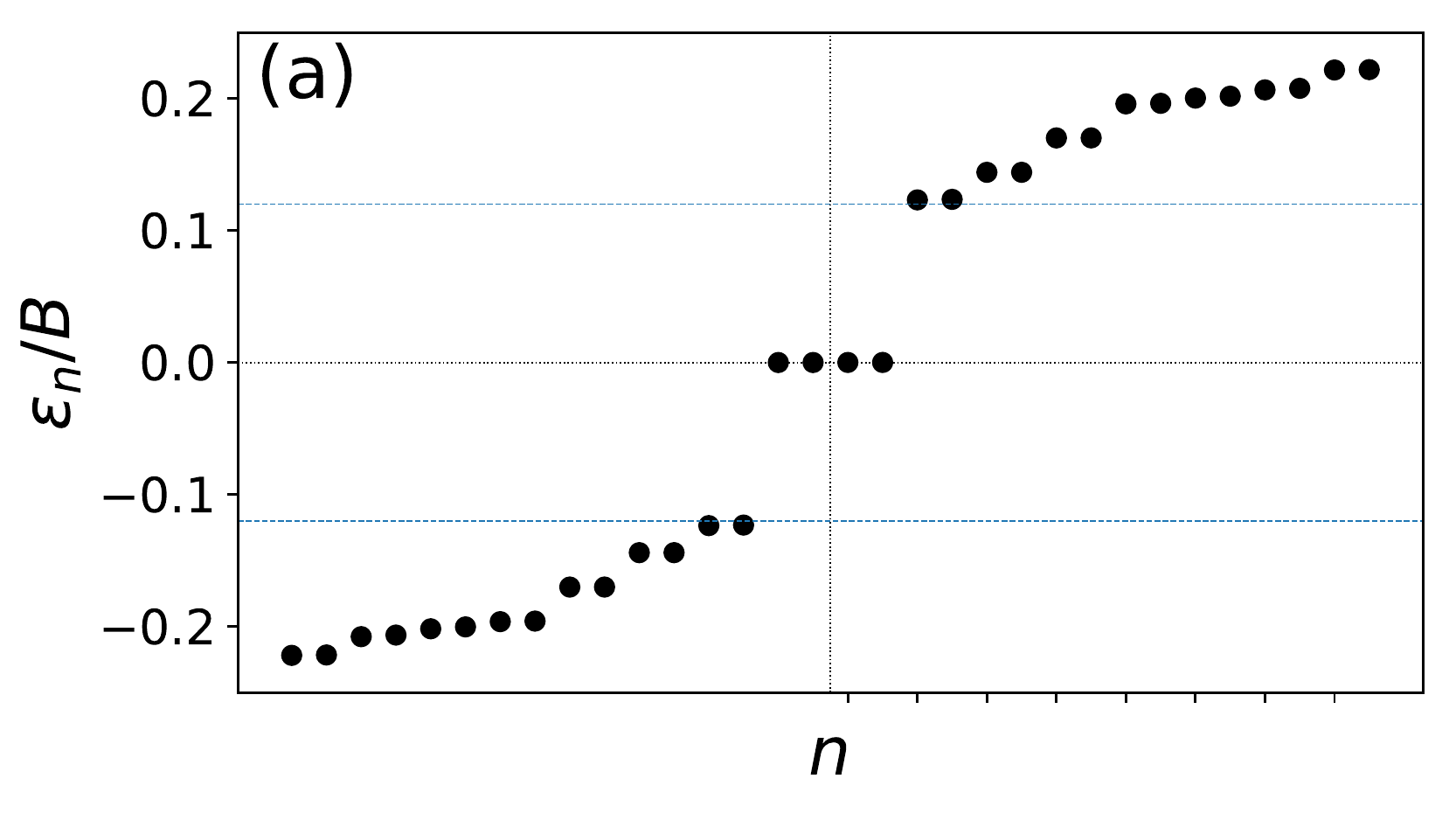}\\
	\includegraphics[width=0.4\textwidth]{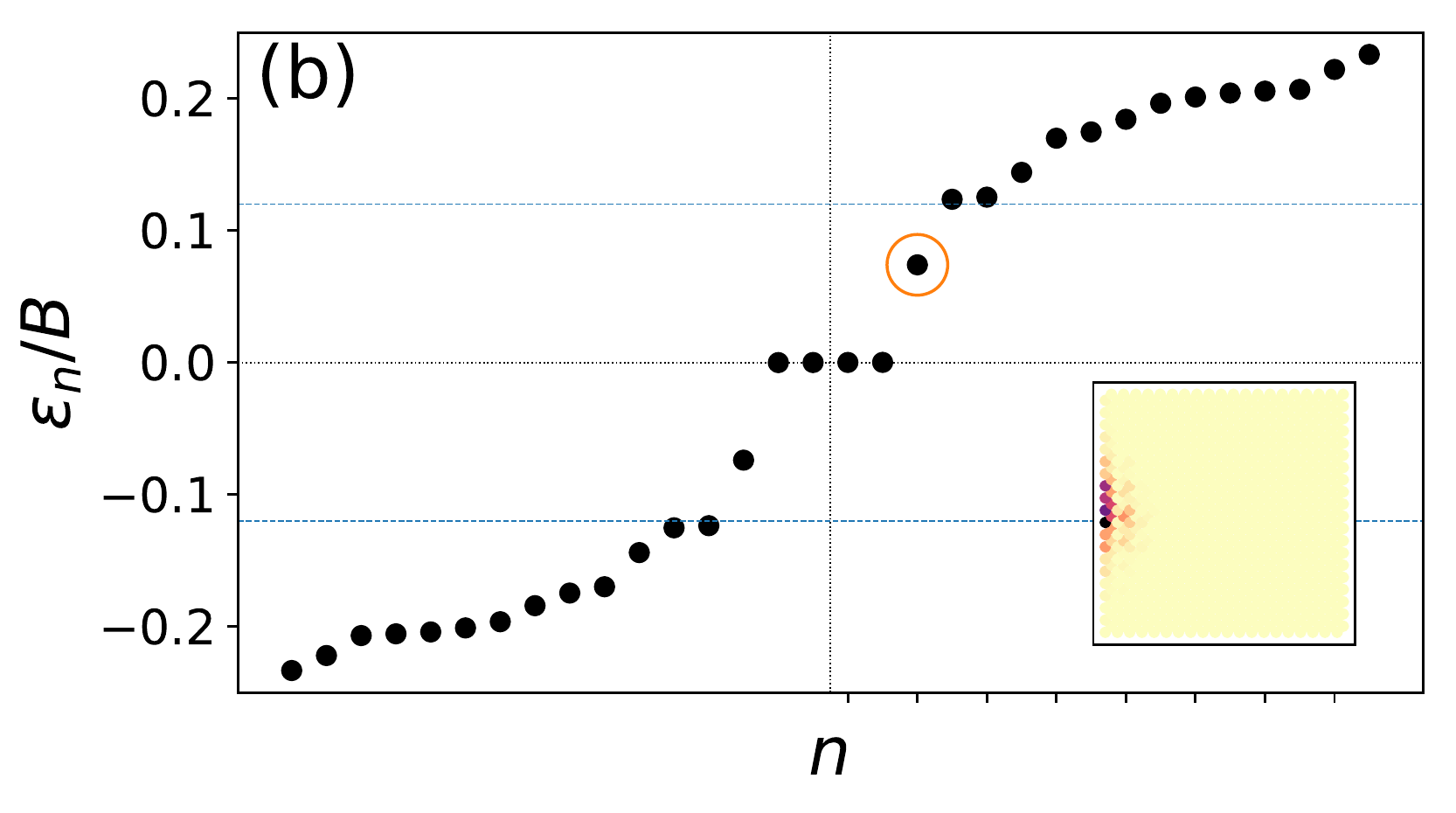}\\
	\includegraphics[width=0.4\textwidth]{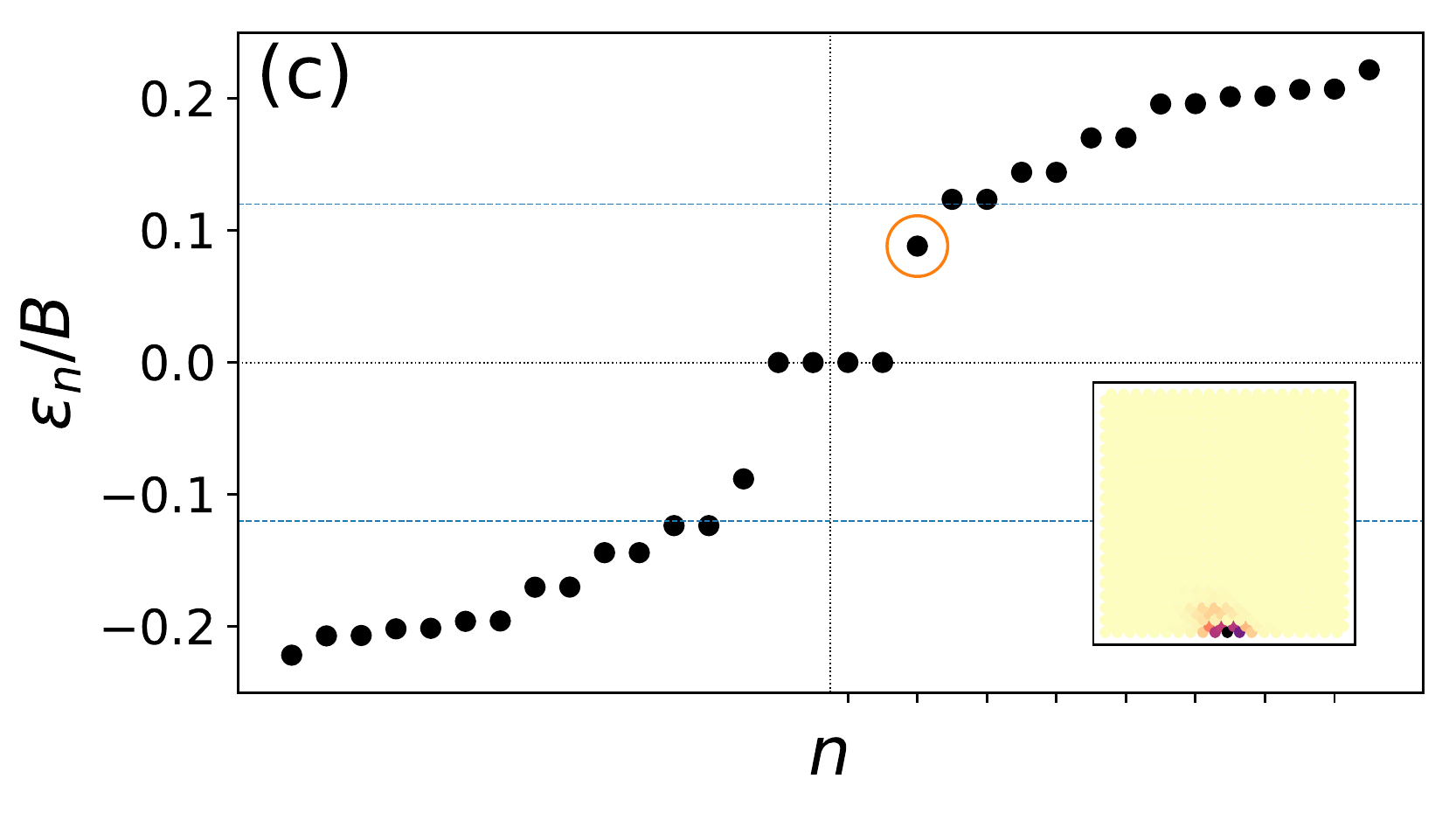}
	\caption{Spectra of $H_{\rm FTS}$ with a single spin-flip impurity located (a) inside the bulk, (b) along an AFM edge, and (c) along an FM edge. In (a) the clean edge gap persists with no additional low-energy bound states; in contrast for (b) and (c) a single sub-edge-gap bound state accompanies the impurity, and the spatial profile of the bound state is shown in inset.
		\label{fig:SI_dis2}}
\end{figure}

This motivates us to consider the effect of individual impurities based on their spatial location. In Fig.~\ref{fig:SI_dis2} we plot spectra for a single (a) bulk, (b) AFM edge, and (c) FM edge impurity. We observe that magnetic impurities (in the form of imperfect bicollinear ordering) do not result in bound states in the bulk below the edge gaps as expected. However, if the impurities are located on the edge, then sub-edge-gap bound states appear (shown in insets). We also see in Fig.~\ref{fig:SI_dis2} that these individual bound states have no apparent impact, however, on the Majorana manifold.

\begin{figure}[t]
	\centering
	\includegraphics[width=0.4\textwidth]{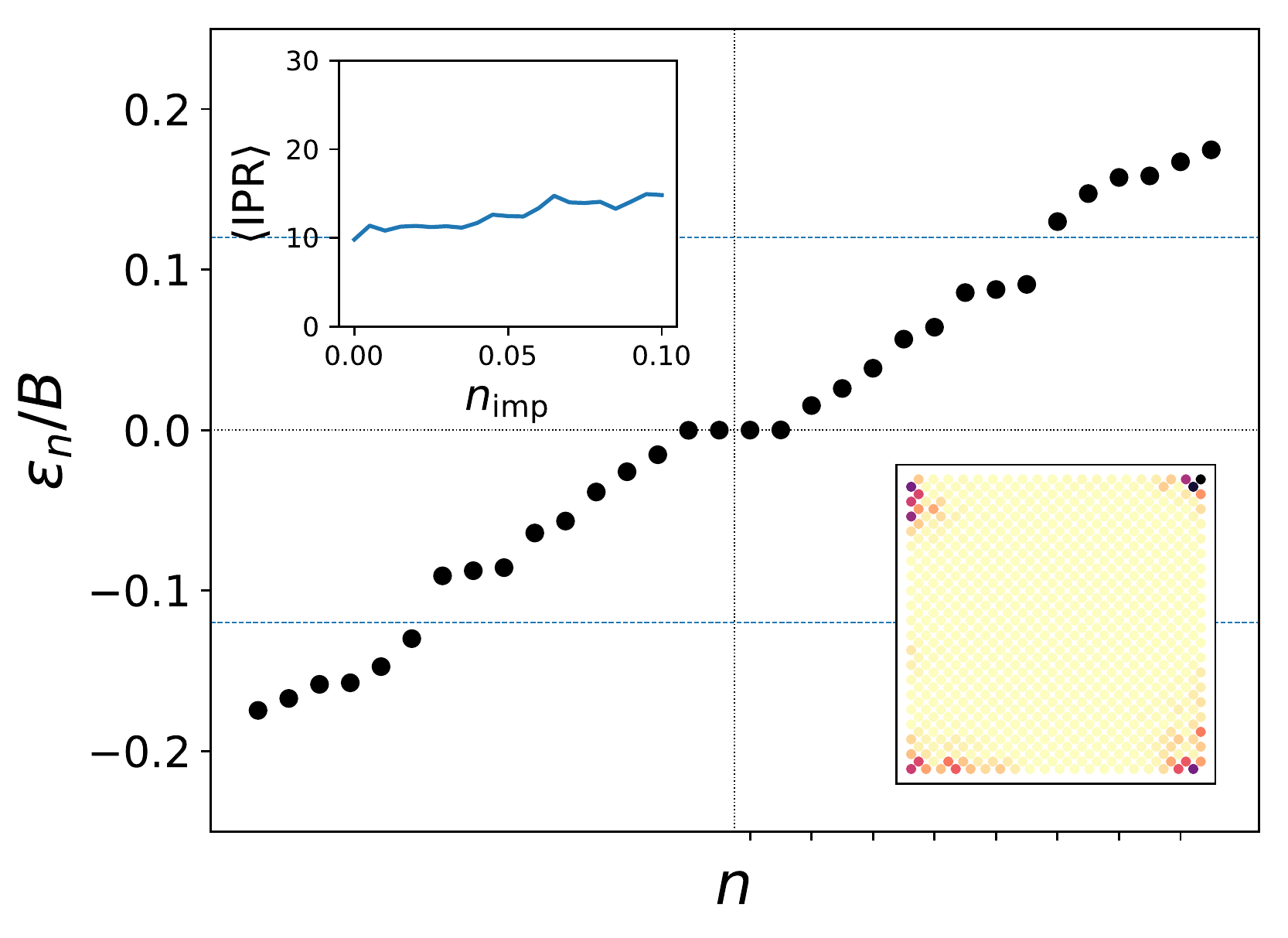}
	\caption{Spectrum of a typical ``gapless'' disorder realization, in the sense that the energy window below the clean edge gap is nearly uniformly filled with states, at $n_{\rm imp} = 0.1$. Nonetheless, the zero modes persist with only slightly decreased localization, as seen from the probability density plotted in the lower-right inset. The upper-left inset shows the disorder averaged IPR for the zero modes, a quantitative measure of localization; up to at least $n_{\rm imp} = 0.1$, increasing impurity density only slightly increases the localization length of the zero modes.
		\label{fig:SI_dis3}}
\end{figure}

We investigate this stability of the Majorana states further by quantitatively characterizing the localization of corner modes through their disorder-averaged inverse participation ratio (averaged also over the four Majorana corner states and the eight particle-hole $\otimes$ orbital $\otimes$ spin basis states),
\begin{equation}
{\rm IPR} = \frac{1}{4 \times 8}\sum_{n \in {\rm MZMs}}\left( \sum_{x,\tau \sigma s} \left| \psi_n(x,\tau \sigma s) \right|^4 \right)^{-1}.
\end{equation}
We show in an inset to Fig.~\ref{fig:SI_dis3} that even as the disorder averaged gap decreases rapidly with increasing $n_{\rm imp}$, the averaged IPR - defined to roughly count the number of sites where the wavefunction is nonzero - is essentially flat, corresponding to a negligible effect of any additional low-energy edge bound states on the corner Majoranas. In Fig.~\ref{fig:SI_dis3} we also show results for a single typical ``gapless'' realization ($n_{\rm imp} = 0.1$) with constant density of states inside the edge gap. Since these subgap states are also localized, they cannot simultaneously hybridize with two corner modes to trivialize them; correspondingly, the combined probability density of the Majorana modes (shown in an inset) still has four disjoint regions of support, but with slightly decreased localization to the corners.

In summary, we have demonstrated that the HOTSC phase we uncover does not rely on perfect bicollinear AFM order: even for relatively large $n_{\rm imp}$ the bulk remains essentially inert (since the bulk QSH does not require the AFM texture). The main effect of increasing $n_{\rm imp}$ then is increasing the likelihood that a particular realization will contain edge impurities. These edge impurities bind additional low-energy localized states, which can hybridize with the Majorana corner modes to ``push" the zero-mode away from the corners, but an isolated zero-mode cannot be eliminated. Instead, a high density of edge impurities is required to push Majoranas from two separate corners together to trivialize them.

\end{document}